\documentclass[letterpaper,twocolumn,10pt]{article}
\usepackage{usenix-2020-09}

\usepackage{tikz}
\usepackage{amsmath}
\usepackage{booktabs}
\usepackage{makecell}
\usepackage{algorithm}
\usepackage{algpseudocode}
\usepackage{multirow}
\usepackage{tabularx}
\usepackage{listings}
\usepackage{appendix}
\usepackage{xspace}
\usepackage{colortbl}
\usepackage{subcaption}
\usepackage{gensymb}
\usepackage{textcomp}
\usepackage{times}
\usepackage{latexsym}
\usepackage{pifont}
\usepackage[inline]{enumitem}
\usepackage{hyperref}
\usepackage{graphicx}
\usepackage{booktabs} 
\usepackage{threeparttable}
\usepackage{microtype}
\usepackage{xparse}

\newcommand{\sysname}{{{TPL-Benchmark}}\xspace}

\usepackage{filecontents}
\usepackage{listings}

\DeclareRobustCommand{\newtexttt}[1]{\lstinline|#1|}

\title{\Large \bf Revisiting Third-Party Library Detection: A Ground Truth Dataset and Its Implications Across Security Tasks}

\author{\rm Jintao Gu\textsuperscript{1*}, Haolang Lu\textsuperscript{1*}, Guoshun Nan\textsuperscript{1}, Yihan Lin\textsuperscript{1}, \\
\rm Kun Wang\textsuperscript{2}, Yuchun Guo\textsuperscript{1}, Yigui Cao\textsuperscript{1}, Yang Liu\textsuperscript{2} \\
\textsuperscript{1}Beijing University of Posts and Telecommunications, China \\
\{3212353402, lhl\_2507, nanguo2021, linjhs, guoyuchun, 2386783626\}@bupt.edu.cn \\
\textsuperscript{2}Nanyang Technological University, Singapore \\
\{wang.kun, yangliu\}@ntu.edu.sg \\
\textsuperscript{*}\rm{Co-first authors}
}

\begin{document}

\maketitle

\begin{abstract}
Accurate detection of third-party libraries (TPLs) is fundamental to Android security, supporting vulnerability tracking, malware detection, and supply chain auditing.
Despite many proposed tools, their real-world effectiveness remains unclear.
We present the first large-scale empirical study of ten state-of-the-art TPL detection techniques across over 6,000 apps, enabled by a new ground truth dataset with precise version-level annotations for both remote and local dependencies.
Our evaluation exposes tool fragility to R8-era transformations, weak version discrimination, inaccurate correspondence of candidate libraries, difficulty in generalizing similarity thresholds, and prohibitive runtime/memory overheads at scale.
Beyond tool assessment, we further analyze how TPLs shape downstream tasks, including vulnerability analysis, malware detection, secret leakage assessment, and LLM-based evaluation. 
From this perspective, our study provides concrete insights into how TPL characteristics affect these tasks and informs future improvements in security analysis.
\end{abstract}
\section{Introduction}

Third-party libraries, which are estimated to account for 60\% to 80\% of the code in typical Android applications~\cite{whatsoftwarecompositionanalysissca2023}, have become an indispensable component of modern Android development ecosystems~\cite{wang2019understanding,wang2018beyond}.
Accurately identifying these libraries, known as TPL detection, is a foundational step in understanding and securing Android applications~\cite{bhoraskar2014brahmastra,enck2011study}.
As a core part of the Android software analysis pipeline, accurate TPL detection enables a precise understanding of application composition~\cite{zhao2023a,jiang2024binaryai}, behavior~\cite{zhou2022uncovering,umayya2024comex,zhang2018re,dong2024same}, and potential risk~\cite{duan2017,li2023malwukong,zheng2024towards}.
It also plays an important role in enabling downstream tasks such as vulnerability identification~\cite{xie2023a,xu2024enhancing,wu2024identifying}, malware detection~\cite{aafer2013droidapiminer,he2022msdroid}, code clone analysis~\cite{wang2015a,li2019identifying}, and repackaging detection~\cite{zhan2019comparative,li2019rebooting}.
As the Android ecosystem continues to grow in complexity, TPL detection has become significantly more challenging due to widespread use of obfuscation, code shrinking, and intricate compilation configurations.
Inaccurate or incomplete detection can, in turn, severely undermine the effectiveness of downstream security analysis.

Over the past decade, extensive research has advanced TPL detection for Android applications significantly.
Early approaches introduced whitelist-based techniques~\cite{li2016,li2019revisiting,samhi2024} that match known library package names to identify imported third-party components, providing a practical starting point and laying the groundwork for tool development.
Subsequent studies explored clustering-based detection techniques~\cite{ma2016,li2017c,zhang2020a}, leveraging structural and semantic features, such as Android API calls, control flow graphs, and opcode patterns, to adapt to the effects of shallow obfuscation techniques and avoid the overhead of maintaining large signature databases. This shift greatly improves scalability and usability, simplifying integration and use in key downstream tasks~\cite{he2022,he2022msdroid,li2025ui}.
To support version-level detection, another line of work proposed similarity-based methods that compute hashed class dependencies and apply fuzzy matching to align apps with version-specific TPL fingerprint databases~\cite{wang2018c,backes2016,zhang2018b,zhang2019,zhan2021a,wu2023a,huang2023a}.
Most recently, researchers have proposed fine-grained static analysis techniques that approximate code transformation and optimization processes~\cite{xie2024b}, making detection more resilient to major code optimization and obfuscation in the modern R8 compiler~\cite{enableappoptimizationappquality}.

\begin{table}[t]
\centering
\caption{TPL Detection Tools Published in Top Venues}
\label{tab:tool_table}
\scriptsize
\setlength{\tabcolsep}{3pt}  
\begin{tabular}{@{}>{\centering\arraybackslash}p{2.2cm}cccccc@{}}
\toprule
\textbf{Tool} & \textbf{M.$^1$} & \textbf{\makecell[c]{Jour.\\/Conf.}} & \textbf{\makecell[c]{Source\\Code\\Open}} & \textbf{\makecell[c]{Dataset\\Reachable}} & \textbf{\makecell[c]{Dataset\\Available}} \\ 
\midrule
AdDetect\cite{narayanan2014} & - & ISSNIP & \ding{55} & \ding{55} & \ding{55} \\ 
WuKong\cite{wang2015a} & C & ISSTA & \ding{55} & \ding{55} & \ding{55} \\ 
Libsift\cite{soh2016} & - & APSEC & \ding{55} & \ding{55} & \ding{55} \\ 
Li et al\cite{li2016} & W & SANER & \ding{51} & \ding{51} & \ding{55} \\ 
Libradar\cite{ma2016} & C & ICSE & \ding{51} & \ding{55} & \ding{55} \\ 
LibScout\cite{backes2016} & S & CCS & \ding{51} & \ding{55} & \ding{55} \\ 
LibPecker\cite{zhang2018b} & S & SANER & \ding{51} & \ding{55} & \ding{55} \\ 
LibD\cite{li2017c} & C & ICSE & \ding{51} & \ding{51} & \ding{55} \\ 
OSSPolice\cite{duan2017} & S & CCS & \ding{51} & \ding{55} & \ding{55} \\ 
LibDetect\cite{glanz2017} & S & ESEC/FSE & \ding{51} & \ding{51} & \ding{55} \\ 
LibScout(extend)\cite{derr2017} & S & CCS & \ding{51} & \ding{55} & \ding{55} \\ 
Han et al\cite{han2018} & S & WPC & \ding{55} & \ding{55} & \ding{55} \\ 
Orlis\cite{wang2018c} & S & ICSE & \ding{51} & \ding{51} & \ding{51} \\ 
Li et al(extend)\cite{li2019revisiting} & W & J.Syst.Softw. & \ding{51} & \ding{51} & \ding{55} \\ 
LibID\cite{zhang2019} & S & ISSTA & \ding{51} & \ding{55} & \ding{55} \\ 
PanGuard\cite{tang2019} & S & CS & \ding{55} & \ding{55} & \ding{55} \\ 
Libseeker\cite{huang2019} & S & JCCS & \ding{55} & \ding{55} & \ding{55} \\ 
LibD(extend)\cite{li2018large} & C & TSE & \ding{51} & \ding{51} & \ding{55} \\ 
LibDX\cite{tang2020a} & S & SANER & \ding{55} & \ding{55} & \ding{55} \\ 
LibExtractor\cite{zhang2020a} & C & WiSec & \ding{55} & \ding{55} & \ding{55} \\ 
LibRoad\cite{xu2020} & S & TMC & \ding{51} & \ding{55} & \ding{55} \\ 
ATVHunter\cite{zhan2021a} & S & ICSE & \ding{55} & \ding{51} & \ding{51} \\ 
AndroLibZoo\cite{samhi2024} & W & MSR & \ding{51} & \ding{51} & \ding{55} \\ 
LibPass\cite{xujian2023} & S & J. Softw. & \ding{51} & \ding{51} & \ding{55} \\ 
PHunter\cite{xie2023a} & S & ISSTA & \ding{55} & \ding{55} & \ding{55} \\ 
LibScan\cite{wu2023a} & S & USENIX Sec. & \ding{51} & \ding{51} & \ding{51} \\ 
LibLOOM\cite{huang2023a} & S & TSE & \ding{51} & \ding{51} & \ding{51} \\ 
LibHunter\cite{xie2024b} & S & ASE & \ding{51} & \ding{51} & \ding{51} \\ 
\bottomrule
\end{tabular}
\begin{minipage}{0.44\textwidth}
\scriptsize
\textit{$^1$ \textbf{M.}: Detection Method (W: Whitelist, C: Clustering, S: Similarity Comparison).}
\end{minipage}
\label{tab:tool_table}
\end{table}

Although existing studies (summarized in Table~\ref{tab:tool_table}) have made valuable contributions to TPL detection, we observe that current approaches still fall short of addressing the complexities of real-world Android applications.
A key limitation is that most methods are evaluated on authors' self-constructed datasets---this introduces bias and restricts the generalizability of results. 
To clarify this issue, we categorized existing dataset construction methods into four main types: \textit{(i)} relying on third-party services~\cite{appbrain}, \textit{(ii)} manual annotation~\cite{wang2018c,wu2023a,huang2023a,zhan2021a}, \textit{(iii)} heuristic-based automation~\cite{li2017c,zhang2018b,li2018large}, and \textit{(iv)} synthetic application construction~\cite{xujian2023,xu2020,zhang2019}.
Each method has contributed meaningfully to the field but faces critical drawbacks: high manual cost, low labeling accuracy, poor scalability, or lack of real-world representativeness.
Collectively, these limitations create inconsistent evaluation standards across studies, impeding fair technique comparison and slowing TPL detection research progress.

To fill this gap, a large-scale, high-quality, representative, and reproducible benchmark dataset is essential for enabling fair and comprehensive evaluation of TPL detection techniques.
Building upon such a dataset, we further aim to address three key research questions that are critical for advancing the field.
With such a dataset in place, we would be able to take this opportunity to examine TPL detection from a broader and deeper perspective, moving beyond tool evaluation to explore its technical limitations and practical implications.
Specifically, this study is guided by the following three research questions:
\begin{itemize} [leftmargin=*, topsep=0pt, itemsep=0pt] 
    \item [\ding{231}] \textbf{From a data construction perspective}, how can we build a trustworthy ground truth for TPL usage in real-world Android applications, at scale and with version-level accuracy?
    \item [\ding{231}] \textbf{From a methodological standpoint}, how do state-of-the-art TPL detection tools perform under a standardized and realistic benchmark, and what limitations do they reveal?
    \item [\ding{231}] \textbf{From a practical perspective}, how does accurate TPL detection contribute to downstream Android analysis tasks, such as vulnerability detection and malware analysis?
\end{itemize}

In this work, we construct \sysname, the largest open-source, real-world dataset for Android TPL detection to date, through a carefully designed pipeline for data collection, annotation, and verification.
We first collected 15,310 candidate repositories from two primary sources (GitHub and F-Droid~\cite{fdroidfreeopensourceandroidapprepository}) and applied a rigorous filtering process to retain 5,876 high-quality project pairs with complete source code.
Next, we applied a semi-automated labeling procedure consisting of \textit{(i)} extracting local and global variables from multiple types of project files, \textit{(ii)} constructing submodule dependency graphs via topological sorting, and \textit{(iii)} identifying the main module to generate reliable triplet labels in the form of \textless Group ID, Artifact ID, Version\textgreater.
Finally, we performed manual-assisted annotation and validation for special or ambiguous cases, followed by sampled manual compilation to further ensure labeling reliability, resulting in a dataset of 6,055 applications with verified TPL labels.

To assess existing techniques' practical effectiveness, we comprehensively evaluated ten state-of-the-art TPL detection methods on \sysname.
This required overcoming key engineering challenges: adapting each tool to a common runtime environment and addressing high computational costs.
By enforcing a unified evaluation protocol, we ensured that the results were directly comparable and representative of realistic Android software scenarios.
The findings reveal notable limitations: at the library level, the highest F1-score achieved was only 60.15\%, whereas the lowest was a mere 2.26\%.
Such low precision and recall, combined with detection times of several hours, suggest that current TPL detection methods remain far from meeting practical needs.
Based on these observations, we performed a targeted analysis of performance bottlenecks and outlined multiple avenues for improvement.

To further demonstrate the practical value of TPL detection, we explored its role in four representative downstream tasks: vulnerability detection, malware discovery, source code secret leakage identification~\cite{zhou2025hey}, and code security auditing.
We identified 3,309 potential TPL--CVE mappings and 8,115 potential App--CVE mappings, showing that vulnerabilities are widely propagated through TPLs and accurate detection is crucial for auditing.
After TPL stripping, 60–80\% of benign code and 40\% of malicious code can be removed, demonstrating that TPL stripping reduces analysis noise while exposing malicious cores.
For source code secret leakage, we found 835,250 secrets across 22 types, highlighting credential exposure as a systemic risk in Android open-source projects.
LLM-based evaluations of 4,500+ projects reveal useful code-level insights yet misalign with real security risks, underscoring both their emerging value and the need for integration with traditional analyses.

\noindent\textbf{Contributions.} This work makes the following contributions:
\begin{enumerate}[leftmargin=*, topsep=0pt, itemsep=0pt]
    \item A systematic pipeline for collecting, annotating, and verifying real-world Android TPL data.
    \item \textbf{\sysname}: the largest open-source benchmark with version-level labels and vulnerability mappings.
    \item A unified evaluation of ten state-of-the-art TPL detection methods, revealing key performance limitations.
    \item Demonstration of TPL detection's value across multiple downstream security tasks.
\end{enumerate}

\section{Background and Prior Knowledge}

\subsection{Android Build System and Dependency Management}
Gradle\cite{gradlebuildtool2024} is the dominant Android build tool and default in Android Studio, with a robust dependency management system that allows developers to declare and manage TPLs directly in build scripts. 
Other tools, such as Maven\cite{mavenrepositorysearchbrowseexplore}, Ant, Buck, Bazel, and CMake, are used far less frequently and typically only in specific scenarios. 
In Gradle-based projects, dependencies (declared in Groovy or Kotlin DSL) may be specified in the project-level \texttt{build.gradle(.kts)} file, which configures plugin versions and global settings.
Module-level build files define the dependencies required throughout the build lifecycle, including compilation, testing, debugging, and runtime; however, not all of these dependencies are packaged into the final APK.
Since our objective is to detect TPLs present in the packaged artifact, we focus on dependencies actually included in the APK, typically declared with keywords such as \texttt{implementation}, \texttt{api}, and others.

\subsection{Import Methods for Third-Party Libraries}
TPLs can be integrated into Android projects in three ways. 
The most common method is through remote repositories, where dependencies are retrieved from sources such as Maven Central or JCenter and uniquely identified by the \textless Group ID, Artifact ID, Version\textgreater\ triplet (e.g., \newtexttt{implementation 'com.android.support:appcompat-v7:28.0.0'}). 
Alternatively, developers may include local binary files, such as \texttt{.jar}, \texttt{.aar}, or \texttt{.so}, directly within the project structure, often under directories like \texttt{libs} or \texttt{deps}. 
A third approach is to import the library's source code as a submodule, which allows direct modification and integration into the build process.

\subsection{Foundations of TPL Labeling}
In this work, we adopt the \textless Group ID, Artifact ID, Version\textgreater\ triplet as the standard identifier for TPLs, enabling consistent labeling across projects. 
Remote repository imports are straightforward to standardize, as their metadata is explicit. 
In contrast, local and submodule imports lack uniform formatting, may contain modified code, and often omit version information, making automatic mapping significantly more difficult. 
Nevertheless, labeling these imports remains valuable, as they increase dataset coverage and capture the diversity of real-world dependency usage, which is essential for building a representative benchmark for TPL detection.

\section{The Proposed \sysname}

\begin{figure*}[htbp]
\centering 
\includegraphics[width=1\textwidth]{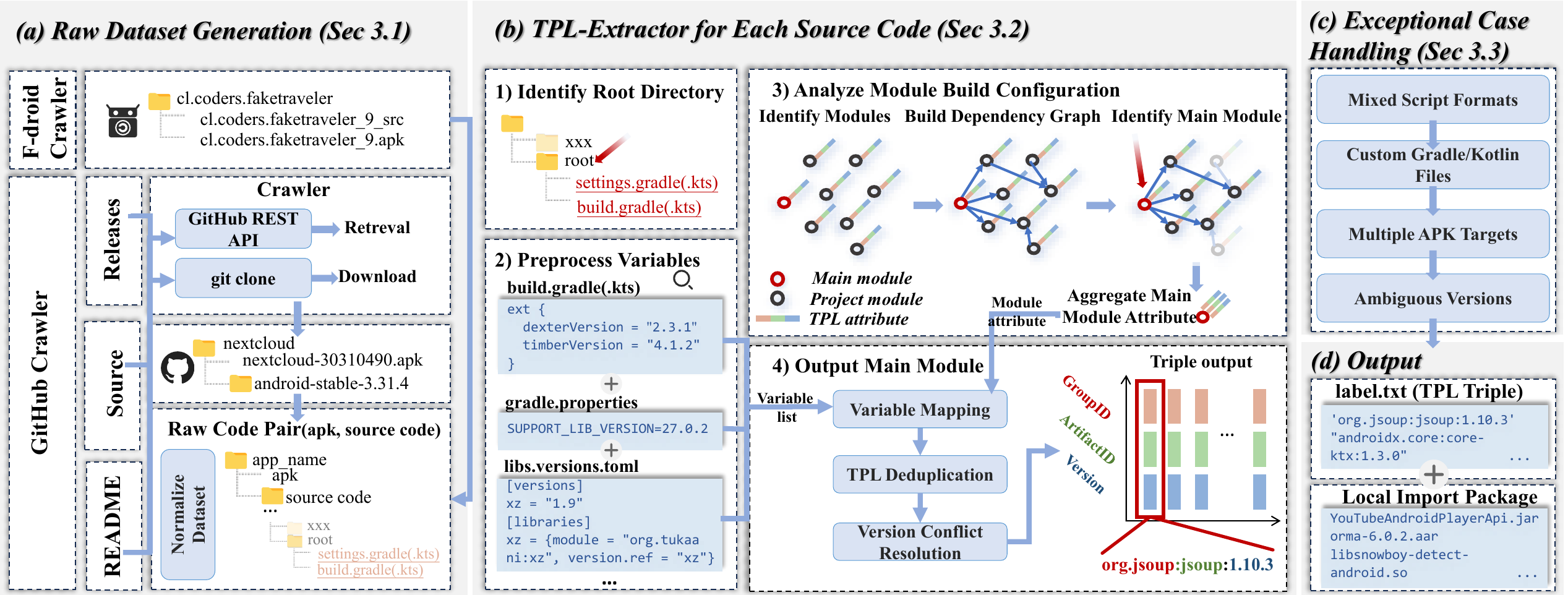}
\caption{Overview architecture. Our pipeline integrates APK-source code pairs, processes them with \textit{TPL-Extractor} to handle complex Gradle configurations, and applies manual validation to ensure accurate and standardized dependency triplets.}
\label{fig:overview}
\end{figure*}

In this section, we present the implementation details of our dataset construction process, organized into four subsections. 
The first three subsections correspond to the \textit{Raw Dataset Generation}, \textit{TPL-Extractor}, and \textit{Exceptional Case Handling} stages illustrated in Figure~\ref{fig:overview}, respectively, detailing how each component operates in practice. 
The final subsection compares the resulting dataset, \sysname, with existing datasets and provides further analysis.

\subsection{Raw data Generation}

To build a large-scale and representative benchmark for TPL detection in real-world Android applications, we collected APK-source code pairs from \textbf{four complementary sources}: F-Droid, GitHub releases containing APKs, GitHub repositories containing APKs in their source code, and repositories whose README files link to APKs.
The collection and filtering procedures for each source are detailed below.

\textbf{F-Droid.}
We retrieved the latest application versions from F-Droid. 
After removing duplicates across categories and excluding projects built with non-Gradle frameworks (e.g., Flutter~\cite{flutterbuildappsanyscreen}, Cordova~\cite{apachecordova}, React Native~\cite{reactnative}), we obtained \textbf{3,551} high-quality Gradle-based pairs.

\textbf{GitHub Releases with APKs.}
We merged two candidate lists: an existing list of 7,862 repositories~\cite{geiger2018} and an additional 2,132 repositories identified via targeted GitHub searches \texttt{(language:java, topic:app)}. 
Using the GitHub REST API, we identified APKs in the latest releases and verified the presence of \texttt{build.gradle(.kts)} files, retaining only Gradle-based projects. 
This process yielded \textbf{962} repositories, some containing multiple APKs.

\textbf{GitHub Source with APKs.}
This category refers to repositories that may include generated APKs within their source code.
We verified version consistency by ensuring that each APK and \texttt{build.gradle(.kts)} file originated from the same commit.
Projects using hybrid build tools or lacking Gradle metadata were excluded, resulting in \textbf{817} eligible repositories, also with some containing multiple APKs (see Appendix~\ref{sec:A} for reasons).

\textbf{README with APKs.}
For the remaining repositories, we checked if their README files contained Google Play links (most of which were invalid) and excluded those with F-Droid links. 
After filtering, we identified 1,634 repositories. We then used APKPure~\cite{apkpure}, a mirror site of Google Play, to locate the corresponding applications and performed manual verification. 
For projects with multiple source-APK versions, we selected only one for matching. 
To ensure version consistency, we extracted the \newtexttt{versionName} from \texttt{AndroidManifest.xml} or build files (e.g., \newtexttt{build.gradle}, \newtexttt{gradle.properties}) and cross-referenced it with repository tags or branches.
Finally, \textbf{547} repositories were retained after screening.

\subsection{TPL-Extractor}

With the curated dataset in place, the next step is to automatically extract reliable TPL labels in the standard \textless Group ID, Artifact ID, Version\textgreater{} triplet form. 
To achieve this, we develop \textit{TPL-Extractor}, a specialized tool that parses diverse Gradle build configurations, resolves dependencies, and normalizes them into standardized labels. 
This section is organized into two parts:  
\textit{(i)} \textbf{Adapting TPL Extraction to Diverse Android Build Configurations}, which examines the key compatibility challenges posed by the heterogeneity of Android build environments;
\textit{(ii)} \textbf{Design and Workflow of \textit{TPL-Extractor}}, which details the architecture and end-to-end processing pipeline we implement to address these challenges.

\subsubsection{Adapting to Diverse Build Configurations}
\label{sec:3.2.1}

Real-world Android projects exhibit substantial heterogeneity in Gradle configurations, dependency declaration styles, and version management practices, which significantly complicates automated TPL extraction. 
An analysis of our dataset shows that a large fraction of projects deviate from default Gradle conventions, requiring targeted handling to extract reliable \textless Group ID, Artifact ID, Version\textgreater{} triplets. 
Below, we highlight the major challenges and the strategies implemented in \textit{TPL-Extractor} to address them.

\textbf{Diverse dependency keywords.}
Gradle supports numerous keywords for declaring dependencies, but not all lead to code being packaged into the final APK.  
Keywords such as \texttt{compile}, \texttt{implementation}, and \texttt{api} typically correspond to packaged code and must be included in TPL detection.  
Others, including \texttt{testImplementation}, \texttt{compileOnly}, and \texttt{debugImplementation}, are used solely for testing, debugging, or compilation phases and should be excluded to avoid false positives.  
Moreover, dynamically defined keywords (e.g., \texttt{nightlyImplementation}, \texttt{gplayImplementation}) vary in inclusion behavior depending on \texttt{flavorDimensions}, \texttt{buildType}, or \texttt{productFlavors}(details in Appendix~\ref{sec:B}).

\textbf{Submodule declaration.} Multi-module projects often declare submodules using patterns such as \textit{(i)} direct declarations (e.g., \newtexttt{compile project(':ExternalSources')} or \newtexttt{api project(':libraries:ColorPickerPreference')}), \textit{(ii)} path-based declarations (e.g., \newtexttt{implementation project(path: ':pages:madani')}), and \textit{(iii)} Kotlin DSL declarations (e.g., \newtexttt{implementation(projects.core.ui)}).

Typically, submodule names can be directly mapped to their corresponding folder paths. However, discrepancies may arise when custom path mappings are defined in the \newtexttt{settings.gradle(.kts)} file, which must be parsed to resolve the correct locations. \textit{TPL-Extractor} also accounts for special submodules such as \newtexttt{wearApp} (e.g., \texttt{project(':wear')}), whose TPL dependencies are not packaged into the main module's \texttt{mobile.apk} but instead compiled into a separate Wear OS \texttt{wear.apk}, which is embedded as a resource in \newtexttt{mobile.apk}.

\textbf{Varied dependency declaration formats.}
Dependency specifications in Gradle are far from uniform(Appendix~\ref{sec:C}).  
We observe string notation declarations (\newtexttt{com.android.support:transition:25.1.0}), map notation declarations (\newtexttt{group: 'com.google.code.gson', name: 'gson', version: '2.8.1'}), inline multi-library forms, and dynamic concatenations using variables (e.g., \newtexttt{"\$group:\$name:\$version"}).  
More complex cases involve cross-module references (\newtexttt{\$rootProject.ext.dependencies.androidx}), project property lookups, or structured dependency bundles defined in \texttt{TOML} or custom \texttt{Gradle/Kt} script file.
Additionally, Bill of Materials (BOM) imports (e.g., \newtexttt{platform('com.squareup.okhttp3:okhttp-bom:4.11.0')}) supply versions from the BOM rather than the declaration location, allowing dependent coordinates to omit explicit versions.  
\textit{TPL-Extractor} unifies these heterogeneous formats into normalized triplets via a dedicated parsing layer.

\textbf{Distributed variable definitions.} In many Android projects, dependency-related variables are defined through diverse patterns(Appendix~\ref{sec:D})---including direct assignments, \texttt{ext} blocks, nested data structures, modular variable sets, and multi-layer dependency objects---and are often spread across project-level and module-level \texttt{build.gradle(.kts)}, \texttt{gradle.properties}, \texttt{TOML}, and custom \newtexttt{Gradle/Kt} scripts (e.g., \newtexttt{deps.gradle}, \newtexttt{versions.gradle}, \newtexttt{dependencies.kt}).
Resolving such variables requires recursive parsing of all relevant files, evaluation of Gradle's property inheritance rules, and expansion of nested definitions.
By combining file-type-specific parsers with a variable resolution engine, \textit{TPL-Extractor} recovers complete and accurate dependency information even in heavily modularized projects.

\subsubsection{Design and Workflow of TPL-Extractor} \label{subsubsec: TPL-Extractor}

Given the diverse structures of real-world Android projects, \textit{TPL-Extractor} is designed to produce a precise and version-specific set of TPL labels from heterogeneous build configurations.  
We formalize the extraction process to ensure clarity, reproducibility, and faithful alignment with Gradle's dependency resolution semantics.

\textbf{Analyze Module Build Configuration.} Mathematically, let $P$ denote the set of Android projects in our dataset.
For each project $p \in P$:
\begin{itemize}[leftmargin=*, itemsep=0pt, topsep=0pt]
    \item $M_p$ is the set of \textit{modules} declared in $p$'s build configuration.
    \item $G_p = (M_p, E_p)$ is the \emph{module dependency graph}.
    \item $m_p^\ast \in M_p$ is the \emph{main application module} that produces the final APK.
\end{itemize}

A module may or may not contribute to the final APK.  
We define the \emph{reachable application modules} as:
\begin{equation}
    A_p \;=\; \{\, m \in M_p \mid m \text{ reachable from } m_p^\ast \ \land\  \phi_p(m)\},
\end{equation}
where $\phi_p(m)$ is a predicate indicating whether module $m$'s outputs are packaged into the main APK (e.g., excluding Wear OS modules, code generators, or deprecated modules).

For each module $m \in M_p$, let $\mathcal{D}_{p,m}$ denote the set of raw dependency declarations found in its build configuration files.
We retain only those Gradle configurations that include code in the APK (e.g., \texttt{implementation}, \texttt{api}), and exclude test-only or debug-only configurations (\texttt{testImplementation}, \texttt{debugImplementation}, etc.).

Dependency declarations may reference variables, properties, or external catalog definitions. We preprocess variables to ensure their reliable resolution. Formally, we define:
\begin{itemize}[leftmargin=*, itemsep=0pt]
    \item $K_p$ is the \textit{ project-level variable bindings} 
    \item $K_{p,m}$ is the\textit{ module-level variable bindings}.
\end{itemize}

Specifically, $K_p$ is extracted from files such as \newtexttt{gradle.properties}, \texttt{TOML}, or project-level \newtexttt{build.gradle(.kts)}, while $K_{p,m}$ is extracted from module-level build scripts.
The \emph{effective environment} for module $m$ is:
\begin{equation}
    V_{p,m} = K_p \oplus K_{p,m},
\end{equation}
where $\oplus$ applies module-level bindings with precedence over project-level bindings.

\textbf{Aggregate Main Module Attribute.}
Rather than resolving dependencies module-by-module, \textit{TPL-Extractor} first aggregates all reachable declarations:
\begin{equation}
    \mathcal{D}_{p,m^*} \;=\; \{\, (m, d) \mid m \in A_p,\ d \in \mathcal{D}_{p,m} \,\}.
\end{equation}
Each element carries its originating module $m$ so that the correct environment $V_{p,m}$ can be applied later.

\textbf{Variable Mapping.}
We define a normalization operator $\mathcal{N}$ that converts raw declarations into explicit \emph{triplets} $(g,a,v)$, where $g$ is the Group ID, $a$ the Artifact ID, and $v$ the Version.
The operator:
\begin{equation}
T^{\mathrm{raw}}_p = \mathcal{N}\big(\mathcal{D}_{p,m^*},\, K_p,\, \{K_{p,m}\}_{m \in A_p}\big),
\end{equation}
performs:
\begin{enumerate}[leftmargin=1.5em, itemsep=0pt, topsep=0pt]
    \item Variable substitution using $V_{p,m}$ for each $(m,d) \in \mathcal{D}_{p,m^*}$.
    \item Expansion of BOM (\emph{Bill of Materials}) references into explicit versions.
    \item Syntax unification across Groovy/Kotlin DSL forms, attribute-based declarations, inline lists, \texttt{TOML} catalogs, and custom Gradle/Kotlin scripts.
\end{enumerate}
The result $T^{\mathrm{raw}}_p$ is a \emph{multiset} of triplets, as duplicates and conflicts may still exist.

\textbf{TPL Deduplication.}  
When aggregating dependencies from all reachable modules, it is common for the same TPL to appear multiple times with identical coordinates $(g,a,v)$, for example, when both the main module and one or more submodules reference the same library version.  
Such duplicates add no new information but unnecessarily inflate the dependency list.  
To address this, we define a deduplication operator $\Delta(\cdot)$ that removes exact duplicates from the raw triplet set:
\begin{equation}
T^{\mathrm{uniq}}_p = \Delta\big(T^{\mathrm{raw}}_p\big),
\end{equation}
where $T^{\mathrm{raw}}_p$ is the union of all normalized dependencies from reachable modules in project $p$, and $T^{\mathrm{uniq}}_p$ is its deduplicated form.

\textbf{Version Conflict Resolution.}  
Even after deduplication, multiple modules in the same project may depend on the same $(g,a)$ coordinate but different versions, leading to version conflicts.  
Gradle resolves such conflicts by selecting a single version for each $(g,a)$ pair (by default, the highest version) unless an explicit override (e.g., \texttt{resolutionStrategy.force}) is provided in the build script.  
We model this process with the conflict resolution function $\rho(\cdot)$:
\begin{equation}
L_p = \rho\big(T^{\mathrm{uniq}}_p\big),
\end{equation}
where $L_p$ is the final, conflict-free set of triplets used for labeling project $p$.

Putting all steps together, the overall extraction process for a project $p$ can be expressed as:
\begin{equation}
    \mathcal{E}(p) = \rho\!\big(\Delta\big(\mathcal{N}(\mathcal{D}_{p,m^*},\, K_p,\, \{K_{p,m}\}_{m\in A_p})\big)\big),
\end{equation}
\begin{equation}
\mathcal{D}_{p,m^*} = \{(m,d) \mid m\in A_p,\ d\in\mathcal{D}_{p,m}\}.
\end{equation}
In practice, \textit{TPL-Extractor} follows the procedure outlined in Algorithm~\ref{alg:tpl_extractor} (See Appendix~\ref{sec:E}), which implements the stages defined in the preceding mathematical formulation.

\subsection{Exceptional Case Handling} 
While the mapping $\mathcal{E}(p)$ captures most dependency configurations automatically, certain irregular cases require manual resolution to ensure that the final label set $L_p$ is correct. We encountered four recurring scenarios:

\textbf{Mixed Script Formats ($K_p \cup \{K_{p,m}\}$).}
Some projects define dependencies in both Groovy (\texttt{build.gradle}) and Kotlin DSL (\texttt{build.gradle.kts}) files. Since automated parsing assumes a single format per module $m \in M_p$, we manually merge the variable and dependency sets before applying normalization $\mathcal{N}(\cdot)$.

\textbf{Custom Gradle/Kotlin Files ($K^{\mathrm{custom}}_p$).}
Irregularly named scripts (See Appendix~\ref{sec:F} for details) may contain variable declarations or direct triplets $(g,a,v)$ outside the standard files considered in $\mathcal{D}_{p,m}$. Automated discovery cannot guarantee completeness, so we manually inspect $K^{\mathrm{custom}}_p$ to extract relevant definitions and integrate them into $\mathcal{D}_{p,m^*}$.

\textbf{Multiple APK Targets ($A^{\mathrm{multi}}_p$).}
If a repository produces multiple APKs, automated construction of $A_p$ cannot reliably determine which APK corresponds to the primary output. Through manual inspection of build files, we determine the correct main module $m_{\mathrm{main}}$ and its associated application, ensuring aggregated dependencies accurately constitute $\mathcal{D}_{p,m^*}$.

\textbf{Ambiguous Versions ($T^{\mathrm{amb}}_p$).}
Open-ended version specifiers (e.g., \texttt{+}, \texttt{latest.release}) complicate resolution in $\mathcal{N}(\cdot)$. Where possible, we consult \texttt{.iml} project files to recover fixed version numbers. Since \texttt{.iml} omits group IDs, collisions between triplets with identical $(a,v)$ but different $g$ remain possible, reinforcing the necessity of triplet-based identification throughout $\mathcal{E}(p)$.

\subsection{Overview of Our \sysname}

With the extraction pipeline formalized in Section~\ref{subsubsec: TPL-Extractor}, each Android project $p$ is ultimately mapped to a set of normalized TPL triplets $L_p$ \textless Group ID, Artifact ID, Version\textgreater.  
Collectively, these labeled pairs $(p, L_p)$ constitute our benchmark dataset, denoted as \sysname.  

In this section, we first clarify the format and semantics of \sysname, directly grounded in the notation introduced earlier.  
We then conduct a systematic analysis along three axes:  
\textit{(i)}~the reliability of our construction methodology;  
\textit{(ii)}~a quantitative comparison of \sysname's scale against existing TPL datasets; and  
\textit{(iii)}~the distributional characteristics of \sysname to assess its coverage of diverse dependency patterns in real-world Android applications.

\subsubsection{Reliability of the Construction Methodology}

To rigorously evaluate the consistency between our extracted dependency annotations $L_p$ and the golden truth $L_p^{\mathrm{gold}}$, we first construct the latter by executing Gradle's built-in dependency analysis for the main application module (e.g., \texttt{app}), targeting the \texttt{releaseRuntimeClasspath} configuration. This configuration captures the complete and resolved set of dependencies required for the application's runtime execution in release mode, as determined by the build tool. Using $L_p^{\mathrm{gold}}$ thus obtained, we employ three complementary quantitative metrics---correlation, absolute deviation, and exact match---to provide a multi-faceted assessment of the reliability of $L_p$.

\textbf{Pearson correlation coefficient.}   
The Pearson correlation coefficient measures the strength of the linear relationship~\cite{cohen2013applied} between the number of TPLs extracted by our method and by compilation:
\begin{equation}
    r = \frac{\sum_{p=1}^n \big(|L_p| - \overline{|L|}\big)\,\big(|L_p^{\mathrm{gold}}| - \overline{|L^{\mathrm{gold}}|}\big)}
{\sqrt{\sum_{p=1}^n \big(|L_p| - \overline{|L|}\big)^2}\,\sqrt{\sum_{p=1}^n \big(|L_p^{\mathrm{gold}}| - \overline{|L^{\mathrm{gold}}|}\big)^2}},
\end{equation}
where $\overline{|L|}$ and $\overline{|L^{\mathrm{gold}}|}$ are the sample means across all $n$ evaluated projects.  
An $r$ value close to $1$ indicates strong proportional alignment between the extracted and golden counts.

\textbf{Mean Absolute Error (MAE).}  
While correlation captures proportionality, it does not reveal magnitude differences~\cite{willmott2005advantages}. We therefore compute:
\begin{equation}
\mathrm{MAE} = \frac{1}{n} \sum_{p=1}^n \left|\,|L_p| - |L_p^{\mathrm{gold}}|\,\right|,
\end{equation}
which measures the average number of libraries by which our extraction differs from the golden truth.

\textbf{Exact Match Rate.}  
To capture the strictest form of agreement~\cite{rajpurkar2016squad}, we compute:
\begin{equation}
\mathrm{MatchRate} = \frac{1}{n} \sum_{p=1}^n \mathbf{1}\!\left[ L_p = L_p^{\mathrm{gold}} \right],
\end{equation}
where $\mathbf{1}[\cdot]$ is the indicator function that equals $1$ if the extracted label set matches the golden truth, and $0$ otherwise.

\textbf{Results.}  
On a random sample of $n=200$ applications, we obtain $r = 0.99$, $\mathrm{MAE} = 0.75$, and $\mathrm{MatchRate} = 0.50$. These results demonstrate that our extraction procedure yields outputs highly consistent with compilation-based ground truth. The small residual discrepancies are primarily caused by build-configuration-induced dependencies (e.g., \texttt{multiDexEnabled}, \texttt{viewBinding}) and plugin-resolved implicit libraries (e.g., \texttt{org.jetbrains.kotlin.plugin.serialization}), which occur infrequently and have minimal impact on overall dataset fidelity.

\subsubsection{Benchmark Scale and Comparative Analysis}

\begin{figure}[t] 
\centering 
\includegraphics[width=0.47\textwidth]{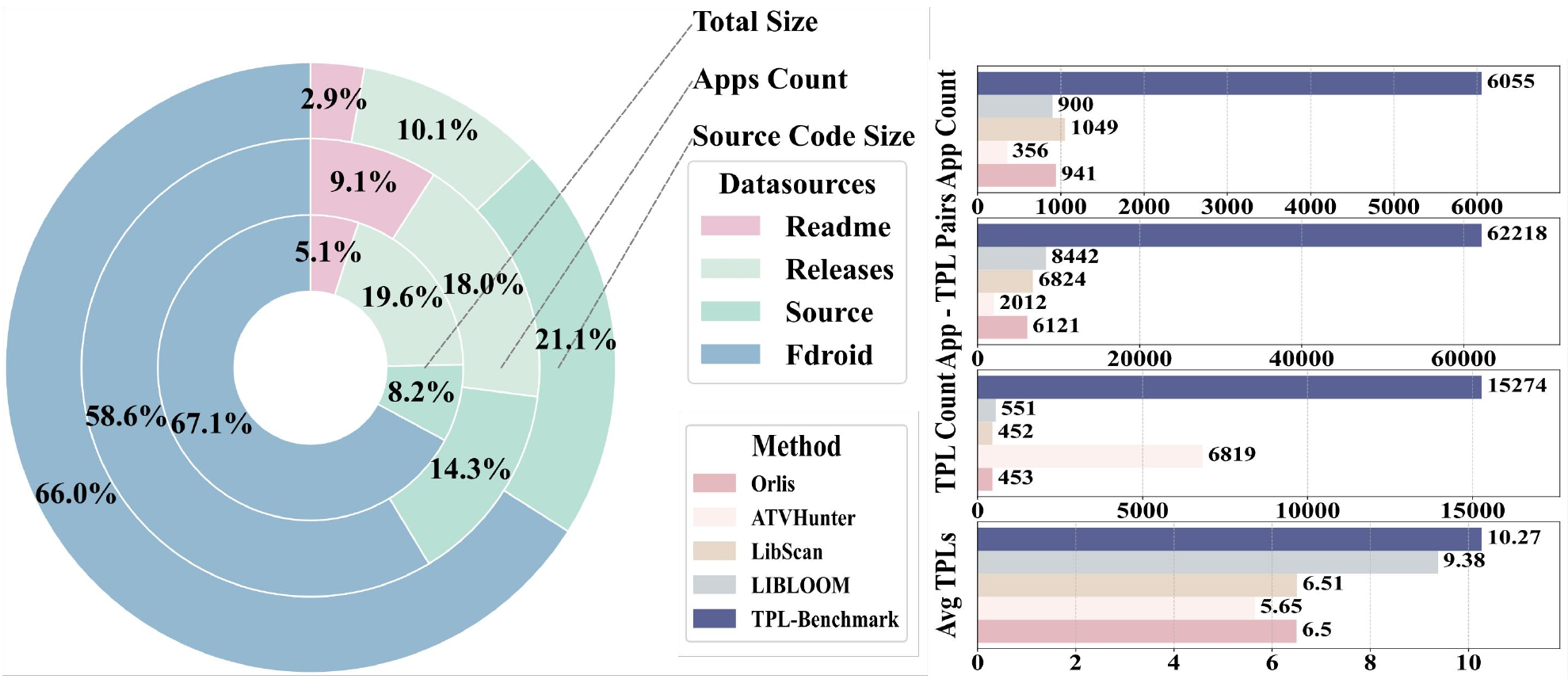} 
\caption{Dataset composition and comparison. The left subfigure shows the statistical data across four distinct data sources. The right subfigure presents an analysis between \sysname and four open-source datasets.}
\label{fig:Statistics_for_Dataset} 
\end{figure}

We compare \sysname against representative open- and closed-source datasets, focusing on those with publicly accessible binaries for reproducibility. Partially available resources (e.g., LibPass, which only provides APK names) are excluded. Among open-source benchmarks, prior works mainly collected apps from F-Droid (e.g., Orlis), extended subsets (e.g., ATVHunter, LibScan), or mixed with proprietary sources (e.g., LibLOOM's blend of F-Droid, ATVHunter, and Xiaomi store apps\cite{xiaomiappstore}). This provides context for scale and diversity comparison.

\sysname achieves substantially higher scale and diversity, with:
\begin{itemize}
    \item \textbf{6,055} real-world apps (8.45 KB--337.28 MB), spanning lightweight to large projects.
    \item \textbf{5,756} unique TPLs and \textbf{15,274} versions.
    \item \textbf{62,218} verified App-TPL mappings.
\end{itemize}

As shown in Figure~\ref{fig:Statistics_for_Dataset}, the average TPL count per app exceeds Orlis, ATVHunter, and LibScan, and slightly surpasses LibLOOM, reflecting broader real-world dependency patterns. Unlike earlier benchmarks dominated by legacy \texttt{com.android.support} libraries, \sysname spans both legacy and modern \texttt{androidx} dependencies, capturing ecosystem evolution. LibLOOM's recent subset shows similar app sizes but narrower library coverage.

\sysname also provides broad \textbf{obfuscation coverage}: over half the apps are built with R8, including shrinking, optimization, and obfuscation by default, reflecting realistic deployment. For evaluating specific obfuscation strategies by different obfuscators~\cite{javaobfuscatorandroidappoptimizerproguard,androidobfuscationjavasecuritydasho2023,allatorijavaobfuscatorprofessionaljavaobfuscation,androidappsecurityobfuscationdexguard,aonzo2020obfuscapk}, controlled datasets (e.g., LibLOOM) remain complementary. Thus, \sysname is well-suited for realistic large-scale benchmarking, while compilation-based synthetic datasets can supplement fine-grained obfuscation tests.

\begin{table}[t]
\centering
\caption{Comparison of Unreachable Method Datasets}
\scriptsize
\begin{tabular}{@{}cccc@{}}
\toprule
\textbf{Method} & \textbf{\begin{tabular}[c]{@{}c@{}}APP\\ Count\end{tabular}} & \textbf{\begin{tabular}[c]{@{}c@{}}TPL\\ Count\end{tabular}} & \textbf{\begin{tabular}[c]{@{}c@{}}App-TPL\\ Paris\end{tabular}} \\ \midrule
WuKong & 200 & \textgreater{}60 & - \\ \cmidrule(l){2-4}
LibSift & 300 & - & - \\ \cmidrule(l){2-4}
AdDetect & 300 & 563 & - \\ \cmidrule(l){2-4}
OSSPolice & - & \begin{tabular}[c]{@{}c@{}}56(native);\\ 279(Java)\end{tabular} & \begin{tabular}[c]{@{}c@{}}295(native);\\ 7055(Java)\end{tabular} \\ \cmidrule(l){2-4}
LibD & 1000 & 2613 & - \\ \cmidrule(l){2-4}
LibDX & 64 & - & 3023 \\ \cmidrule(l){2-4}
LibRoad & \begin{tabular}[c]{@{}c@{}}GT1$^1$:300;\\ GT2:1000\end{tabular} & - & \begin{tabular}[c]{@{}c@{}}GT1:2800;\\ GT2:14233\end{tabular} \\ \cmidrule(l){2-4}
LibPass & \begin{tabular}[c]{@{}c@{}}GT1:10000;\\ GT2:1000;\\ GT3:656\end{tabular} & - & \begin{tabular}[c]{@{}c@{}}GT1:95434;\\ GT2:14233;\\ GT3:3224\end{tabular} \\ \cmidrule(l){2-4}
LibPecker & 9834 & 310 & 33964 \\ \cmidrule(l){2-4}
LibID & \begin{tabular}[c]{@{}c@{}}GT1:1045;\\ GT2:953\end{tabular} & 69(1444$^2$) & - \\ \cmidrule(l){2-4}
\sysname & 6055 & 5756(15274$^2$) & 62218 \\ \bottomrule
\end{tabular}
\begin{minipage}{0.38\textwidth}
\scriptsize
\textit{$^1$ ``GT'' denotes ground truth subsets. $^2$ Including version-level TPLs.}
\end{minipage}
\label{tab:non-open-source}
\end{table}

Beyond open-source datasets, we also compare with non-open benchmarks (Table~\ref{tab:non-open-source}) along three dimensions:
\textit{(i)}~\textbf{Transparency.}  
Some datasets (e.g., LibExtractor, PanGuard, Libradar) do not fully disclose construction procedures, limiting ground-truth confidence. Others (e.g., Han~et~al., LibScout) target narrower goals (malware detection, specific library verification), making them less general-purpose.
\textit{(ii)}~\textbf{Scale and Diversity.}  
WuKong, LibSift, and OSSPolice cover fewer apps. LibID and LibD add more variety but rely on manual compilation or heuristics, reducing reproducibility. LibPecker includes more apps but less library diversity, and heuristic annotations may be obfuscation-sensitive.
\textit{(iii)}~\textbf{Repackaging-based synthetic datasets.}  
Datasets like LibPass and LibRoad inject libraries after removing originals, enabling controlled tests but reducing code realism and execution fidelity.

In contrast, \sysname is built from complete real-world apps, preserving original logic, dependency relationships, and developer practices. Combined with its obfuscation coverage, it offers high-fidelity large-scale benchmarking. When fine-grained obfuscation control is needed, it can be paired with compilation-based synthetic datasets for a comprehensive evaluation suite.

\subsubsection{Manual Intervention for Complex Dependency Configurations}

Beyond overall scale and diversity, it is equally important to understand the internal distributional characteristics of \sysname. 
Such characteristics reveal how TPL usage manifests in practice and provide deeper insights into the challenges faced by detection tools. 

In particular, we focus on three key aspects: 
\textit{(i)} the distribution of library versions, which reflects the evolutionary dynamics of dependencies; 
\textit{(ii)} the prevalence of locally imported libraries, which complicates detection beyond public repositories; 
and \textit{(iii)} the Android Gradle Plugin (AGP) versions employed, which influence build behavior and compatibility. 
Together, these factors highlight the practical complexity captured by \sysname and underscore its value as a benchmark grounded in real-world development practices.

\textbf{Version Distribution.}
Gradle supports multiple version declaration styles. 
In \sysname, we observe five types: 
\textit{(i)} exact version (e.g., \texttt{1.1}); 
\textit{(ii)} range (e.g., \texttt{[1.0,)}); 
\textit{(iii)} prefix (e.g., \texttt{1.1.+}); 
\textit{(iv)} dynamic latest (e.g., \texttt{latest.release}); 
and \textit{(v)} snapshot (e.g., \texttt{1.1-SNAPSHOT}).

\begin{figure}[t] 
\centering 
\includegraphics[width=0.48\textwidth]{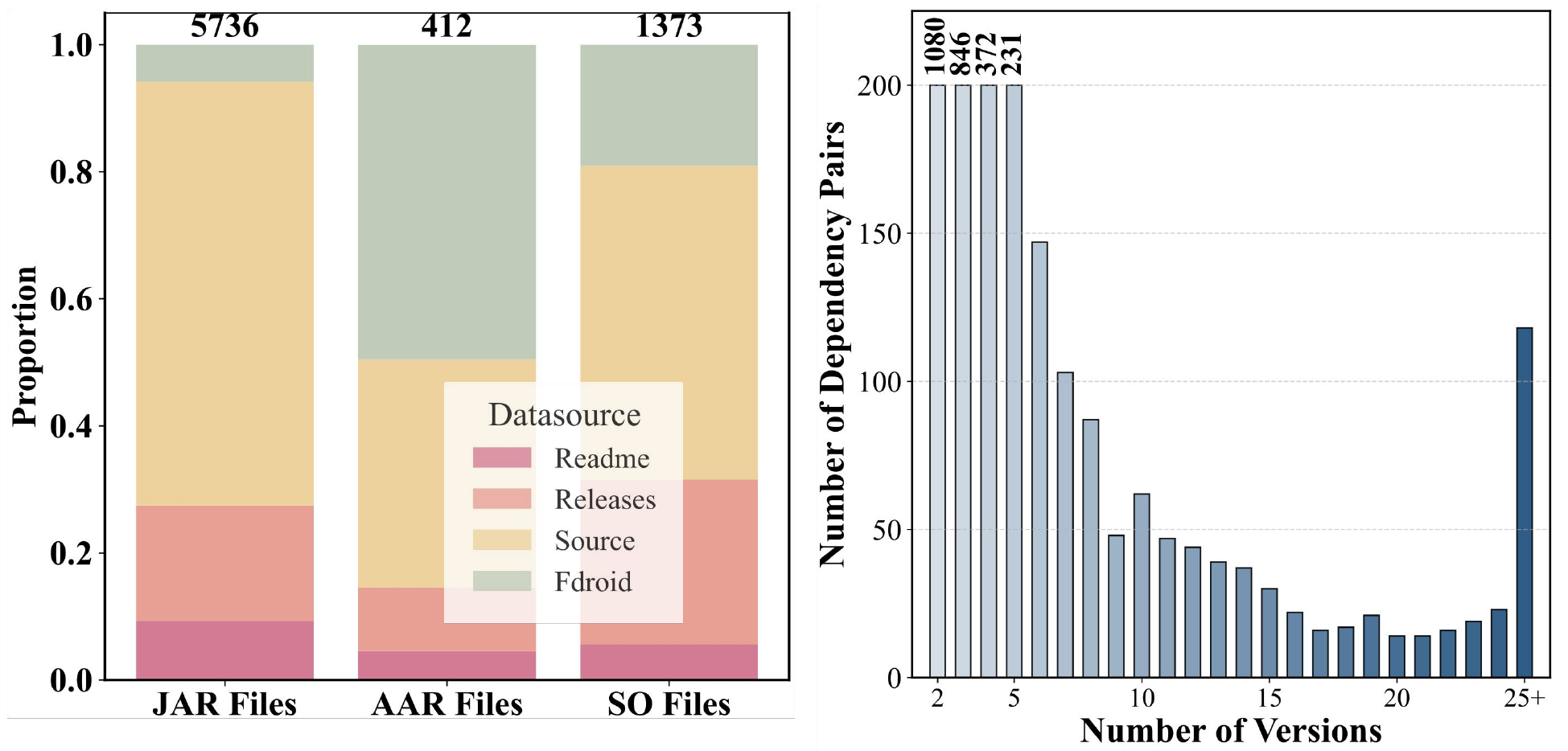} 
\caption{Distribution of Locally Imports and TPL Versions. The left subfigure shows the distribution of locally imported library files (JAR, AAR, SO). The right subfigure illustrates the distribution of versions for TPLs.} 
\label{fig:local_tpl_and_version_distribution} 
\end{figure}

Figure~\ref{fig:local_tpl_and_version_distribution} shows widespread version diversity. 
For example, \texttt{com.google.android.material:material} is represented by 74 distinct versions. 
In total, 1,894 dependencies have more than one version, 748 more than three, 464 more than five, and 237 more than ten.

This diversity reflects library evolution and long-term maintenance challenges, while also exposing security risks from outdated versions. 
Thus, version distribution in \sysname is a valuable lens for studying evolution trends and vulnerability persistence in Android libraries.

\textbf{Local Imports.}
Besides remote repositories, \sysname systematically captures locally imported libraries, including \texttt{.jar}, \texttt{.aar}, and native \texttt{.so} files. 
Although less frequent, these imports remain a non-negligible part of the dependency ecosystem.

As shown in Figure~\ref{fig:local_tpl_and_version_distribution}, the dataset contains 5,736 \texttt{.jar}, 412 \texttt{.aar}, and 1,373 \texttt{.so} files, averaging 0.98, 0.07, and 0.23 per project, respectively. 
Local imports are more common in earlier AGP versions, indicating a historical reliance on manual or offline integration, while modern projects increasingly adopt centralized resolution via remote repositories.

The most prevalent local file is \newtexttt{android-support-v4.jar}, reflecting legacy practices prior to Maven/Gradle repositories. 
Native libraries such as \texttt{libtessellator.so} and \texttt{libpath\_ops.so} appear frequently in graphical processing tasks. 
These findings highlight both the evolution of dependency management and the importance of preserving local imports in comprehensive TPL profiling.

\begin{table}[t]
\centering
\caption{AGP version Statistics for \sysname}
\scriptsize 
\begin{tabular}{@{}cccc@{}}
\toprule
\textbf{AGP version} & \textbf{Interval} & \textbf{Compiler} & \textbf{\begin{tabular}[c]{@{}c@{}}Projects\\ Number\end{tabular}} \\ \midrule
1.0.1-3.1.0 & 2015.1-2018.3 & DX & 2129 \\
3.1.0-3.4.0 & 2018.3-2019.4 & D8 & 524 \\
3.4.0-8.0.0 & 2019.4-2023.4 & R8(Compatibility Mode) & 2131 \\
8.0.0-8.8.0 & 2023.4-2025.1 & R8(Full Mode) & 1039 \\
none & - & - & 53 \\ \bottomrule
\end{tabular}
\label{tab:Compiler_statistics}
\end{table}

\textbf{AGP Version Coverage.}
\sysname spans a broad spectrum of AGP versions and compiler backends, from legacy DX to modern R8 full-mode optimizations (Table~\ref{tab:Compiler_statistics}). 
This distribution captures the evolutionary trajectory of the Android build toolchain, ensuring both historical breadth and present-day relevance.

Such coverage enables two benefits. 
First, it supports longitudinal studies of ecosystem practices, allowing analysis of how library usage, build performance, or obfuscation strategies evolve across generations. 
Second, it facilitates compatibility-aware benchmarking: tools such as decompilers or library detectors can be evaluated under varied compilation outputs, from DX bytecode to aggressively optimized R8 builds.

By including recently compiled projects alongside legacy ones, \sysname reflects both the continuity and evolution of real-world Android development. 
This dual perspective makes it well-suited for studying long-term trends while also enabling up-to-date evaluations under modern toolchains.

\begin{table*}[t]
\centering
\caption{Comprehensive Evaluation of Ten TPL Detection Tools on \sysname}
\scriptsize
\begin{tabular}{@{}ccccccccccccc@{}}
\toprule
\multirowcell{3}[-1.2em][c]{\textbf{Method (Category)}}
 & \multicolumn{4}{c}{\textbf{Profiling Phase}} & \multicolumn{2}{c}{\textbf{Detection Phase}} & \multicolumn{6}{c}{\textbf{Result}} \\ \cmidrule(l){2-13} 
 & \multicolumn{2}{c}{\textbf{Libs}} & \multicolumn{2}{c}{\textbf{Apps}} & \multirowcell{2}[-0.9em][c]{\textbf{\begin{tabular}[c]{@{}c@{}}Success\\ Count$^2$\end{tabular}}} & \multirowcell{2}[-0.9em][c]{\textbf{\begin{tabular}[c]{@{}c@{}}Average\\ Time\end{tabular}}} & \multicolumn{3}{c}{\textbf{Library Level}} & \multicolumn{3}{c}{\textbf{Version Level}} \\ \cmidrule(lr){2-5} \cmidrule(l){8-13} 
 & \textbf{\begin{tabular}[c]{@{}c@{}}Success\\ Count$^1$\end{tabular}} & \textbf{\begin{tabular}[c]{@{}c@{}}Total\\ Time\end{tabular}} & \textbf{\begin{tabular}[c]{@{}c@{}}Success\\ Count$^2$\end{tabular}} & \textbf{\begin{tabular}[c]{@{}c@{}}Total\\ Time\end{tabular}} &  &  & \textbf{RC} & \textbf{PR} & \textbf{F1} & \textbf{RC} & \textbf{PR} & \textbf{F1} \\ \midrule
\multicolumn{13}{l}{\textbf{Clustering-based}} \\ 
Libradar & - & - & - & - & \textbf{946} & \textbf{9.37s} & 9.19\% & 13.58\% & 10.96\% & - & - & - \\
LibD & - & - & - & - & 911 & 447.81s & 41.33\% & 16.24\% & 23.31\% & - & - & - \\ \midrule
\multicolumn{13}{l}{\textbf{Similarity-based (full database)}} \\ 
Orlis & 11442 & 39h57m & - & - &\textbf{946}& 233.28s & 1.22\% & 14.74\% & 2.26\% & 0.79\% & 1.74\% & 1.08\% \\
LibScout & 12043 & 3h10m & - & - & 942 & \underline{38.05s} & 41.07\% & 14.81\% & 21.77\% & 38.54\% & 3.97\% & 7.19\% \\
LibLOOM & 12394 & 33m & 946 & 12m & \textbf{946}& 38.56s & \underline{57.15\%} & 18.81\% & 28.31\% & \underline{50.24\%} & 7.25\% & 12.67\% \\ \midrule
\multicolumn{13}{l}{\textbf{Similarity-based (candidate set)}} \\ 
LibPecker & 12167 & - & - & - & \textbf{946} & 223.13s & 47.11\% & 83.15\% & \textbf{60.15\%} & 42.33\% & 58.96\% & \textbf{49.28\%} \\
LibScan & 12004 & - & - & - &\underline{945} & 43.69s & 13.98\% & \textbf{89.13\%} & 24.18\% & 12.87\% & \underline{60.06\%} & 21.20\% \\
LibID-S & 12167 & 13h19m & 939 & 42h57m & 923 & 3679.61s & 34.30\% & \underline{87.62\%} & 49.30\% & 31.44\% & \textbf{60.41\%} & \underline{41.35\%} \\
LibID-A & 12167 & 13h19m & 939 & 42h57m & 932 & 3977.24s & 33.61\% & 84.32\% & 48.06\% & 30.45\% & 57.27\% & 39.76\% \\
LibHunter & 12004 & - & - & - & 944 & 342.40s & \textbf{81.79\%}& 39.07\% & \underline{52.87\%} & \textbf{76.48\%} & 17.55\% & 39.69\% \\ \bottomrule
\end{tabular}
\begin{minipage}{0.93\textwidth}
\scriptsize
\textit{$^1$ Total Libs Count: 13717. $^2$ Total Apps Count: 946. \textbf{Bold} indicates the largest value in a column; \underline{underline} indicates the second-largest.}
\end{minipage}
\label{tab:combined_evaluation}
\end{table*}

\section{Evaluation and Discussion of TPL Detection}

\subsection{Data Preparation}

To evaluate TPL detection tools fairly, we first constructed a large-scale reference database of 13,940 libraries (jar and aar) collected from Maven, Aliyun, and JitPack repositories over one month. 
For similarity-based methods, we designed two experimental settings: one compares applications against the entire TPLs database, while the other employs a customized subset to balance scalability and efficiency.

Since many tools do not directly support aar files, we batch-converted them into jars, successfully processing 8,814 out of 9,037. 
The 223 failures were mainly resource-only packages (e.g., \texttt{firebase-ads:10.0.0}, \texttt{play-services:10.0.1}), which contain no executable jars. 
In total, we retained 13,717 libraries for inclusion in \sysname. 
From our benchmark, we then selected 946 applications (mostly with $>$20 TPLs) for evaluation. 
After excluding 841 ground-truth jars with little or no code features, we obtained 28,210 valid App-TPL mappings.

\subsection{Efficiency, Compatibility and Effectiveness }

For comparison, we grouped the ten detection methods into three categories: 
\textit{(i)} \textbf{Clustering-based} (Libradar, LibD), which only output package structures rather than specific libraries or versions; 
\textit{(ii)} \textbf{Similarity-based (full database)} (LibScout, LibLOOM, Orlis), which match against the entire library set; 
\textit{(iii)} \textbf{Similarity-based (candidate set)} (LibPecker, LibScan, LibID-S, LibID-A, LibHunter), where we used a controlled candidate strategy including ground-truth libraries, 100 randomly selected other versions of those libraries, and 100 randomly selected libraries not included in the previous two categories to ensure fairness and efficiency (the parameter settings and evaluation details are provided in Appendix~\ref{sec:G}).

We evaluated efficiency (time and success rates in profiling/detection), compatibility (ability to handle diverse apps and libraries), and effectiveness (library-level and version-level), and Table~\ref{tab:combined_evaluation} shows clear differences across tools.

\textbf{Profiling phase.} 
LibLOOM was most practical, constructing 12,394 signatures in 33 minutes (90.4\%). 
LibScout also struck a good balance with 12,043 signatures in 3h10m. 
By contrast, Orlis (39h57m) and LibID (13h19m) achieved similar coverage (88--90\%) but at prohibitive costs, underscoring that only LibLOOM and LibScout scale well to large datasets.

\textbf{Detection phase.} 
LibLOOM, LibScan, LibHunter, LibScout, and LibPecker processed nearly all 946 apps, while LibID-S, LibID-A, and LibD lagged slightly (911--932 apps) due to timeouts or memory limits. 
Orlis and Libradar nominally succeeded on all apps, but only 16.5\% and 72.6\% yielded valid results, showing that raw ``success count'' overstates true utility. 
The minor discrepancy between LibID-S and LibID-A arose from skipping extremely slow cases, not algorithmic inconsistency.

\textbf{Execution overhead.} 
Libradar was fastest (9.37s/app) but too imprecise for reliable use. 
LibLOOM and LibScout maintained strong efficiency (38s/app). 
LibPecker and LibHunter were slower (200--400s/app) but delivered higher precision, suiting quality-focused tasks. 
LibID-S/A were the slowest (over 1h/app, up to 288h), making them impractical at scale but useful for exhaustive forensic analysis.

\textbf{Summary.} 
LibLOOM offers the best balance of efficiency, compatibility, and speed. 
LibPecker achieves the strongest precision-recall trade-off (F1=60.15\%), while LibHunter delivers the highest recall (81.79\% library-level, 76.48\% version-level) with moderate cost, making it promising for large-scale scans. 
LibID variants remain precise but infeasible for deployment. 
Clustering-based tools are lightweight but too coarse for reliable identification.

\subsection{Analysis and Outlook}

Our comparative evaluation demonstrates that existing TPL detection tools differ substantially in efficiency, effectiveness, and practical usability. 
Importantly, these discrepancies are not random but rather stem from several root causes inherent to modern Android app development and build practices.

\textbf{Root causes.} 
First, \textbf{aggressive shrinking, obfuscation, and optimization} significantly disrupt structural features. 
In our dataset, when shrink ratios exceeded 50\%, similarity-based methods such as LibScout and LibPecker consistently exhibited sharp drops in recall, since their threshold-based matching could no longer accommodate heavily pruned method and class signatures. 
Second, libraries belonging to the \textbf{same root package} (e.g., \texttt{butterknife} vs. \texttt{butterknife-runtime}) created namespace ambiguity in candidate selection. 
Tools relying on package-structure heuristics often misattribute code fragments to the wrong candidate, producing systematic false positives. 
Third, current tools almost completely miss \textbf{Kotlin-Specific libraries}. 
For instance, none of the detectors successfully identified TPLs that only contain \texttt{kotlin\_metadata}, confirming a systematic blind spot as Kotlin Multiplatform packages become increasingly prevalent. 
Finally, \textbf{feature sparsity and structural similarity} further complicate detection. 
Optimized lightweight libraries often leave too few features for reliable matching, while structurally similar libraries are easily confused under obfuscation, reducing both precision and recall.
These findings suggest that fundamental methodological limitations, rather than dataset coverage alone, drive the observed gaps.

\textbf{Leveraging Resources and Metadata.} 
Most existing approaches rely almost exclusively on code-level features such as package hierarchies, method signatures, or opcode sequences. 
These features are inherently vulnerable to compiler-driven transformations, including shrinking, obfuscation, and aggressive optimizations. 
Future detection frameworks should complement code-level analysis with resilient non-code artifacts, such as distinctive resource files~\cite{zhauniarovich2014fsquadra,shao2014towards}, configuration metadata, or build-related fingerprints, which tend to persist across compilation and transformation pipelines.

\textbf{Compiler-Aware and AGP-Adaptive Detection.} 
The continuous evolution of the AGP and its compilers (DX $\rightarrow$ D8 $\rightarrow$ R8) introduces increasingly aggressive optimizations~\cite{wu2023a,xie2024b}, with R8 full mode (post-8.0.0) posing unprecedented challenges to static analysis.
Next-generation TPL detection frameworks should adopt compiler-aware strategies that adapt to AGP-version-specific transformation patterns. 
For example, knowledge of enabled passes (e.g., method inlining, class merging) can inform the use of de-optimization heuristics or invariant representations tailored to each compiler version.

\textbf{Adaptive Feature Granularity.} 
Fine-grained features (e.g., instruction-level representations) provide strong discriminative power and enable version-level differentiation, but incur prohibitive computational costs at scale. 
Conversely, coarse-grained methods (e.g., package-level or API-based) achieve efficiency but lack precision, particularly for libraries with similar structural patterns or multi-library coexistence within shared packages.
A promising direction is to design adaptive mechanisms that dynamically select feature resolution based on library variance or preliminary heuristics, balancing precision and scalability.

\textbf{Structural Robustness.} 
Our results show that many representative tools (e.g., LibD, LibScout, LibID) remain highly sensitive to structural perturbations such as package flattening or class reorganization, both of which are common outcomes of R8 optimizations and obfuscation. 
Future research should prioritize the development of structure-invariant representations such as graph embeddings~\cite{xu2017neural} or learned representations from graph neural networks~\cite{yu2020order} that encode semantics beyond raw structure, thus resisting such transformations.

\textbf{Downstream Task Friendliness and Integration.} 
Current TPL detectors typically output a bare list of libraries, offering no readily actionable artifacts (e.g., code boundaries~\cite{li2025ui}, the exact code regions belonging to each library~\cite{he2022msdroid}, stripped application variants~\cite{li2019revisiting}) that downstream tasks could directly consume without additional preprocessing. 
For example, similarity-based methods rarely identify the exact code segments belonging to each detected library, which is crucial for fine-grained analyses such as malware behavior attribution.
Moreover, downstream program analysis tools (e.g., FlowDroid~\cite{arzt2014flowdroid}, IccTA~\cite{li2015iccta}, DroidSafe~\cite{gordon2015information}, Amandroid~\cite{wei2018amandroid}) could benefit significantly from ``library-stripped'' app versions to avoid redundant analysis of common libraries, improving efficiency and success rates.
Future TPL detection frameworks should offer ready-to-use, plug-and-play outputs---similar to Libradar~\cite{ma2016}---allowing downstream tasks to operate directly without rebuilding feature databases or performing complex preprocessing.

\section{Downstream Tasks: Empirical Investigations}
Building on the \sysname dataset, we further investigate how TPL detection connects to broader security and software engineering tasks. 
Specifically, we examine four representative directions: \textit{(i)} the prevalence of vulnerabilities within TPLs, \textit{(ii)} the role of TPL detection in strengthening Android malware analysis, \textit{(iii)}the risk of secret leakage in the Android open source projects, and \textit{(iv)} the feasibility of leveraging LLMs for code understanding. 
Through these explorations, we uncover key insights that highlight both opportunities and limitations of current practices. These findings not only validate the utility of \sysname beyond detection effectiveness, but also suggest concrete avenues for future research in secure and intelligent mobile software analysis.

\subsection{Vulnerabilities in Third-Party Libraries}

Leveraging vulnerability annotations from the National Vulnerability Database (NVD)~\cite{nvdhome}, we established a comprehensive mapping between CVEs and the 15,264 libraries in \sysname. 
In total, 728 libraries contain at least one CVE, producing 3,309 potential TPL--CVE pairs. 
By associating applications with their dependent libraries, we identified 1,607 potentially affected apps (over one-quarter of the dataset), leading to 8,115 App--CVE mappings across 514 unique CVEs. 
This quantifies the breadth of supply-chain risk propagated via TPL usage at scale.

\begin{figure}[t] 
\centering 
\includegraphics[width=0.48\textwidth]{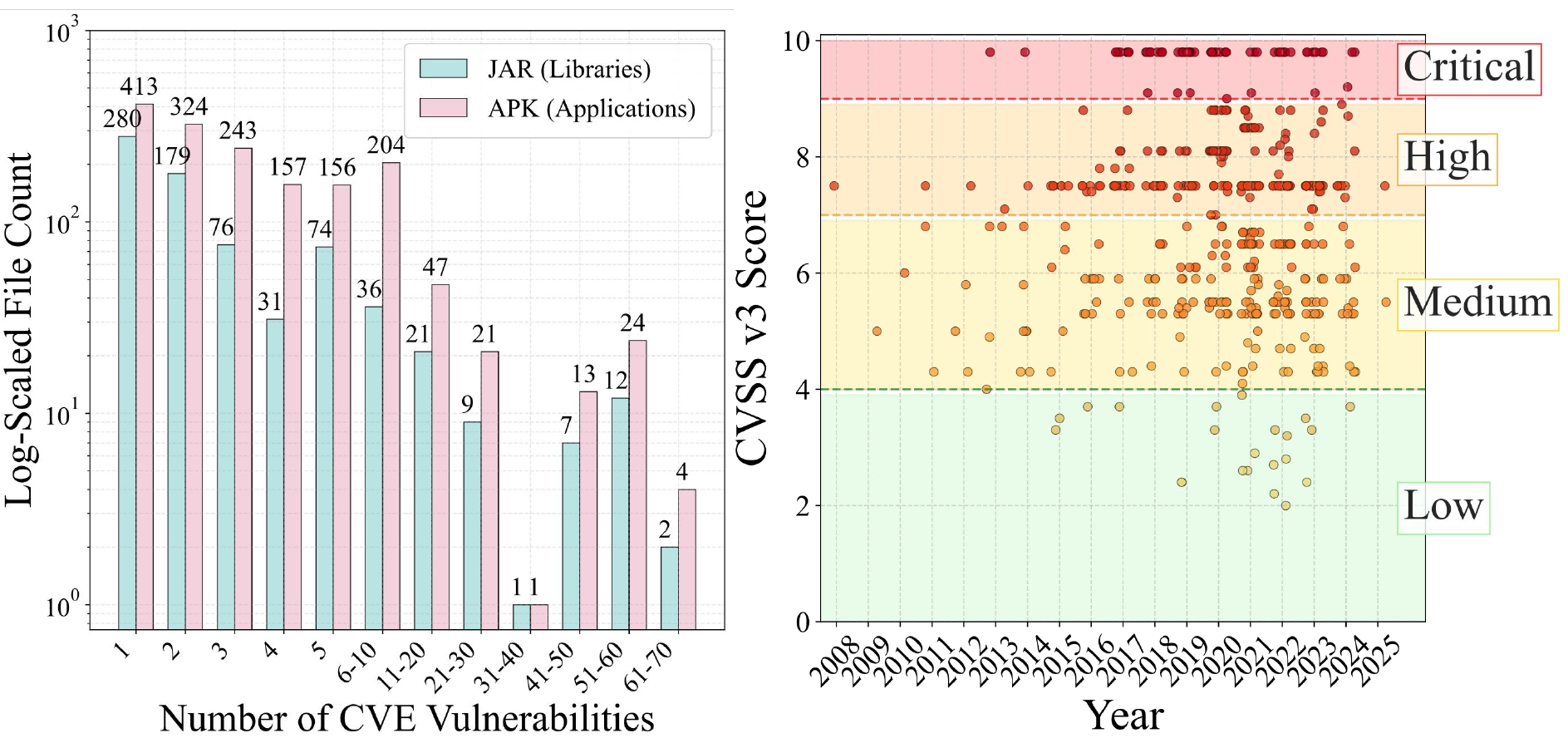} 
\caption{Distribution and Severity of CVEs in TPLs and Apps. The left subfigure shows the number of TPLs and Apps under different CVE counts, and the right subfigure shows the CVSS severity scatter plot of CVEs.} 
\label{fig:apk_jar_cve_distribution} 
\end{figure}

Several libraries and applications exhibit particularly severe cases. 
For instance, \newtexttt{com.fasterxml.jackson.core:jackson-databind:2.9.5} alone is linked to 64 CVEs, while the application \textit{Guide7} is associated with 67 CVEs. 
Distribution analysis in Figure~\ref{fig:apk_jar_cve_distribution} shows that most libraries ($\approx$93\%) contain 1 to 10 CVEs, yet 52 libraries exceed 10 CVEs, indicating a non-trivial subset of highly vulnerable components. 
CVSS scores further reveal that over half of the identified CVEs fall under \emph{High}/\emph{Critical} severity (right panel of Figure~\ref{fig:apk_jar_cve_distribution}), underscoring concrete operational risk.

\textbf{Why these libraries?} A functional breakdown points to systemic exposure in data parsing stacks (e.g., Jackson, XStream), which are naturally prone to deserialization abuse (e.g., CVE-2020-24750, CVE-2021-39147). 
Networking/communication layers (e.g., Netty, OkHttp) and database components (e.g., SQLite) accumulate protocol/IO-bound attack surfaces. 
Even ubiquitous utility suites (e.g., Guava's CVE-2018-10237, Hutool's CVE-2023-24163) demonstrate that ``foundational'' dependencies are not intrinsically safer; their wide reuse amplifies impact when flaws emerge.

\textbf{Version and era effects.} To test whether severity merely reflects legacy code, we examined apps with $\geq$10 CVEs (39 projects) and grouped them by AGP era. 
About 60\% target earlier AGP versions (1.x--4.x), as expected, since older apps ship outdated libraries.
However, nearly 40\% target modern AGP (7.x--8.x), revealing that newer apps still ship with vulnerable, outdated libraries despite patched alternatives being available. 
Two factors likely contribute: \textit{(i)} \textbf{awareness and signaling}. Developers prioritize functionality and may lack timely signals that a dependency has known CVEs or that a patched, compatible version exists; \textit{(ii)} \textbf{compatibility overhead}. Upgrading core libraries often requires resolving non-trivial compatibility issues, which creates inertia even when patched versions are available.

\textbf{Insights.} Beyond counting CVEs, these findings argue for integrating \textbf{version-lag} and \textbf{patch-availability-but-unadopted} metrics into TPL risk assessment. 
For research benchmarks, this suggests complementing CVE coverage with temporal measures of remediation delay. 
For engineering practice, SCA pipelines should surface \emph{actionable replacements} (safe coordinates and minimal-change upgrade paths) and fail the build when high-severity CVEs have readily available fixes. 
In short, \sysname not only captures real-world vulnerable components, but also exposes upgrade inertia as a first-order driver of residual risk, informing both detector evaluation and supply-chain governance.

\subsection{Malware}

The widespread integration of TPLs into Android applications introduces non-trivial obstacles for malware detection. 
Because many libraries (particularly advertising, analytics, and utility SDKs) are equally common in both benign and malicious apps, their extensive code often dominates feature representations while contributing little discriminative value~\cite{avdiienko2015mining,aafer2013droidapiminer}. 
This ``feature dilution'' effect can obscure behavioral signals that are crucial for distinguishing malicious activity. 
To revisit this issue, we curated a balanced dataset from AndroZoo~\cite{Allix:2016:ACM:2901739.2903508,alecci2024androzoo}, including 4,681 benign apps (confirmed benign by all VirusTotal engines) and 1,018 malicious apps (flagged by at least 20 engines) between 2020--2025. 
Only one version was retained to avoid redundancy. 
Using our curated TPL whitelist, we performed static analysis on call graphs to assess the impact of TPL presence and removal.

\begin{figure}[t] 
\centering
\includegraphics[width=0.48\textwidth]{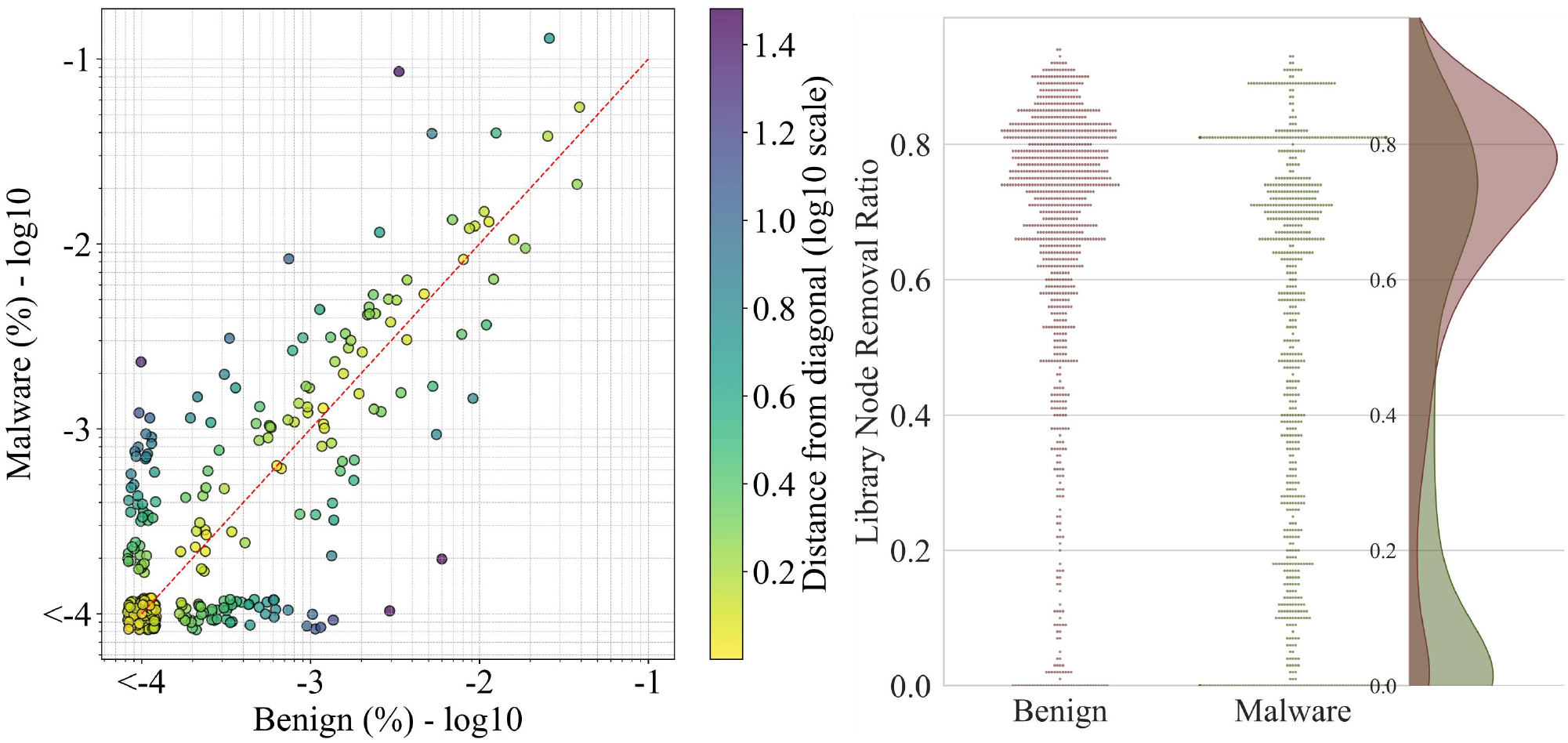}
\caption{TPL Presence in Benign and Malicious Apps. The left subfigure shows the distribution of individual TPLs across benign and malicious APKs, while the right subfigure illustrates the overall TPL removal proportion in apps.}
\label{fig:removal_ratio_violin_box_scatter}
\end{figure}

\textbf{Overall distribution.} 
As shown in Figure~\ref{fig:removal_ratio_violin_box_scatter}(right), benign apps cluster around high TPL recognition ratios (60--80\%), suggesting that their call graphs are heavily dominated by widely used libraries. 
Malicious apps instead exhibit a more uniform distribution, with a significantly larger fraction showing extremely low TPL ratios. 
Upon closer inspection, we find that the call graphs of these malicious apps are typically relatively small, typically containing only tens to hundreds of nodes, and therefore lack any TPLs themselves. 
This contrast suggests that benign apps are often ``library-heavy'', with redundant dependencies that obscure behavioral signals. 
In comparison, a subset of malicious apps concentrates functionality in smaller graphs, where malicious logic constitutes a disproportionately large share.

\textbf{Efficiency implications.} 
To quantify the impact of pruning library nodes, we applied a malware detection pipeline (Malscan~\cite{wu2019malscan}) to both full and TPL-pruned call graphs. 
Even when aggressively retaining only developer nodes and removing up to 70\% of the graph, F1 scores remained stable (86--88\%), while storage requirements dropped from 238GB to 71GB and feature extraction time decreased by 20\%. 
This suggests that the bulk of TPL nodes contribute little to detection accuracy, yet dominate computational overhead---supporting the case for library-aware pruning in malware pipelines.

\textbf{Library-level discriminative power.} 
Beyond global proportions, Figure~\ref{fig:removal_ratio_violin_box_scatter}(left) shows that many TPLs align closely with the diagonal, meaning they appear in similar ratios across benign and malicious samples and thus offer negligible discriminatory power. 
By contrast, libraries far from the diagonal reveal asymmetric tendencies: Google service packages cluster toward benign apps, reflecting mainstream developer reliance on official SDKs, while certain analytics and tracking SDKs (e.g., \texttt{com.yandex.metrica}) are disproportionately present in malicious samples, suggesting potential use in privacy-invasive behavior. 
Interestingly, Kotlin coroutine libraries also appear more frequently in malware, indicating that modern asynchronous frameworks may be leveraged for stealthy execution, dynamic code loading, or evasion.

\textbf{Insights.} 
Together, these findings highlight a dual role of TPLs in malware analysis: ubiquitous libraries in benign apps dilute discriminative signals, while several key libraries can indeed serve as a differentiation aid for malware.
Improved TPL detection accuracy could therefore enable selective node removal strategies that sharpen the semantic contrast between benign and malicious graphs, improving both efficiency and interpretability. 
Ultimately, TPL-aware feature selection appears to be a promising direction for building malware classifiers that are both more robust and more efficient at scale.

\subsection{Secret Leakage}

To assess the risk of secret leakage in Android open-source projects, we applied \texttt{Gitleaks} to all repositories in \sysname. 
As shown in Table~\ref{tab:Statistics_of_Secret_Leakage}, the scan identified a total of \textbf{829{,}580 leaked secrets} across \textbf{682 projects}, corresponding to a leakage prevalence of \textbf{19.2\%}. 
The distribution varied substantially across four channels. F-Droid projects were by far the most affected, with an average of 233.6 leaks per project and a maximum of 151,610 leaks in a single repository. 
By comparison, the Source and Release channels exhibited moderate averages (1.25 and 4.56 leaks per project), though still showed extreme outliers with up to \textbf{3{,}088} and \textbf{391} leaks. 
Even the smaller-scale README channel had confirmed exposures, reflecting insufficient secret management.

\begin{table}[t]
\centering
\caption{Statistics of Secret Leakage}
\scriptsize
\begin{tabular}{@{}ccccccc@{}}
\toprule
\textbf{Channel} & \begin{tabular}[c]{@{}c@{}}\textbf{Leaky}\\ \textbf{Projects}\end{tabular} & \begin{tabular}[c]{@{}c@{}}\textbf{Total}\\ \textbf{Leaks}\end{tabular} & \begin{tabular}[c]{@{}c@{}}\textbf{Prevalence}\end{tabular} & \begin{tabular}[c]{@{}c@{}}\textbf{Avg}\\ \textbf{Leaks}\end{tabular} & \begin{tabular}[c]{@{}c@{}}\textbf{Max}\\ \textbf{Leaks}\end{tabular} & \begin{tabular}[c]{@{}c@{}}\textbf{Std}\\ \textbf{Dev}\end{tabular} \\ \midrule
Readme & 99 & 267 & 18.1\% & 0.49 & 26 & 1.9 \\
Releases & 185 & 4386 & 19.2\% & 4.56 & 3088 & 100.82 \\
Source & 207 & 1017 & 25.4\% & 1.25 & 391 & 13.89 \\
Fdroid & 682 & 829580 & 19.2\% & 233.62 & 151610 & 4208.93 \\ \bottomrule
\end{tabular}
\label{tab:Statistics_of_Secret_Leakage}
\end{table}

\textbf{Content and File-Type Associations.}
From a content perspective, the majority of leaks were generic API keys (803,293) and private keys (26,179), both of which are particularly concerning because they are prone to being packaged into distributed APKs. 
File-type analysis revealed strong structural associations: API keys appeared predominantly in \texttt{.rsp} and \texttt{.txt} files, while private keys were concentrated in \texttt{.json}, \texttt{.txt}, and \texttt{.pem} files, with \texttt{.json} alone contributing nearly 39\%. 
Similarly, PKCS\#12 certificates were almost exclusively leaked in \texttt{.p12} format. 
Particularly alarming, cloud service credentials (e.g., \texttt{GCP} and \texttt{AWS} API keys) were disproportionately exposed in automatically generated artifacts such as \texttt{.snapshot} and \texttt{.dill} files, pointing to build pipelines as an overlooked leakage source.
These \textit{file-type} skews indicate that leakage clusters in configuration-like and build-generated artifacts rather than source files alone.

\textbf{Leakage Mechanisms Along the Build Pipeline.}
The observed distributions suggest several plausible pathways by which secrets enter public repositories and releases.
First, secrets stored in configuration files (\texttt{.json}/\texttt{.txt}/\texttt{.pem}) are often committed directly and persist across tags, which explains their dominance and why archival platforms like F-Droid continue to expose them over time.
Second, build and CI/CD processes can unintentionally propagate credentials into intermediate artifacts. For example, the concentration of PKCS\#12 material in \texttt{.p12} files and the over-representation of cloud keys in \texttt{.snapshot}/\texttt{.dill} outputs indicate that generated caches and artifacts are non-trivial leakage vectors, yet often overlooked by static scanners.
Third, extreme outliers across four channels may stem from auxiliary bundles or logs accidentally included during packaging, which can embed large volumes of sensitive data in a single release.
Taken together, these mechanisms provide a plausible link between the observed ``what/where'' distributions and operational practices along the development--build--release pipeline.

\textbf{Insights.}
The exceptionally high number of secret leaks in some F-Droid projects shows how open archival practices amplify long-term risks: once secrets are committed, historical states preserve them indefinitely, keeping them accessible to adversaries. 
This persistence makes simple pre-release scrubbing insufficient, as stale but valid credentials may already be exposed. 
Therefore, Mitigation requires a \textbf{full-lifecycle perspective}, where detection and prevention span development, build, test, release, and maintenance. 
Practically, CI/CD pipelines should integrate high-risk  \textit{file-type} scanners, while IDE plugins can warn developers prior to commit. 
From a research angle, our dataset enables modeling of \textit{credential lifetime} and \textit{file-type risk}, supporting more precise security policies for Android workflows.

\subsection{LLM-based Code Specification Evaluation}

To systematically investigate whether LLMs can provide meaningful insights into software quality and security, we designed a multi-stage evaluation pipeline using the Qwen-Plus model~\cite{team2024qwen2}. 
Each Android project in our dataset was analyzed across code quality, code security, testing coverage, and documentation/community practices, based on its source code directory. 
A standardized prompting procedure was used, and outputs were normalized into an ordinal A--E scale, with higher weights assigned to security (45\%) and code quality (30\%) as primary indicators. 
To handle input size constraints, we employed a progressive summarization mechanism that extracted bounded file-tree structures and code snippets on demand, enabling the LLM to reason about project characteristics without exceeding its context limitations. 
For each project, the model produced a JSON-based evaluation report, alongside a separate record of potential credential exposures.

\begin{figure}[t] 
\centering 
\includegraphics[width=0.47\textwidth]{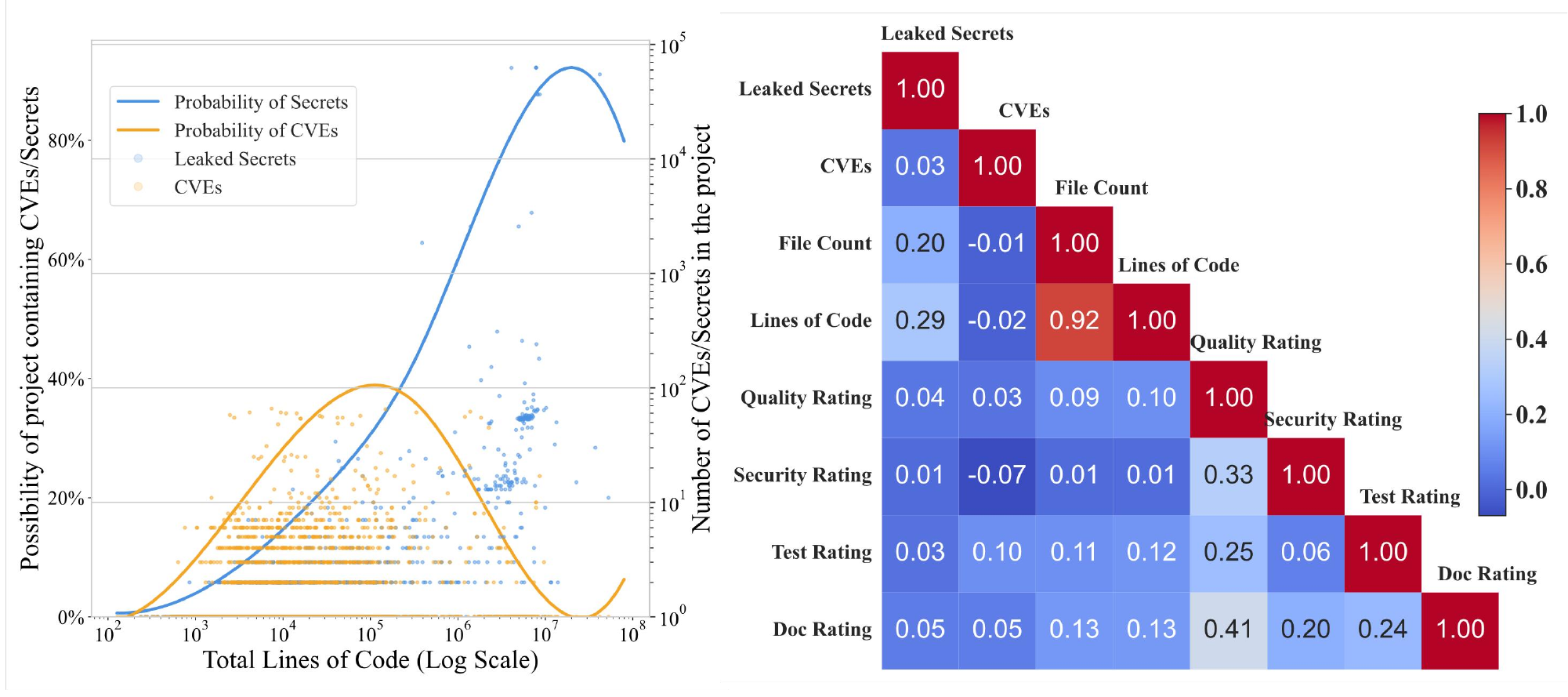} 
\caption{Vulnerability and Secret Exposure Distribution with Correlation Analysis of Project Attributes. The left subfigure presents the distribution of CVE and secret exposures across projects with different lines of code. The right subfigure displays the correlation matrix between LLM evaluation dimensions and project statistics.}
\label{fig:Correlation_Analysis} 
\end{figure}

\textbf{Statistical Correlation Analysis.}  
We examined whether LLM-generated scores aligned with empirical security outcomes. 
As shown in Figure~\ref{fig:Correlation_Analysis}, pearson correlation coefficients between LLM evaluations and indicators such as secret leakage prevalence and CVE labeling revealed only weak associations. 
This suggests that while LLMs are sensitive to surface-level or stylistic cues of software engineering practices, they are not yet reliable predictors of deep semantic vulnerabilities. 
From a research perspective, this points to a fundamental gap between heuristic judgments made by LLMs and ground-truth risks observed in practice.

\begin{figure}[t] 
\centering 
\includegraphics[width=0.47\textwidth]{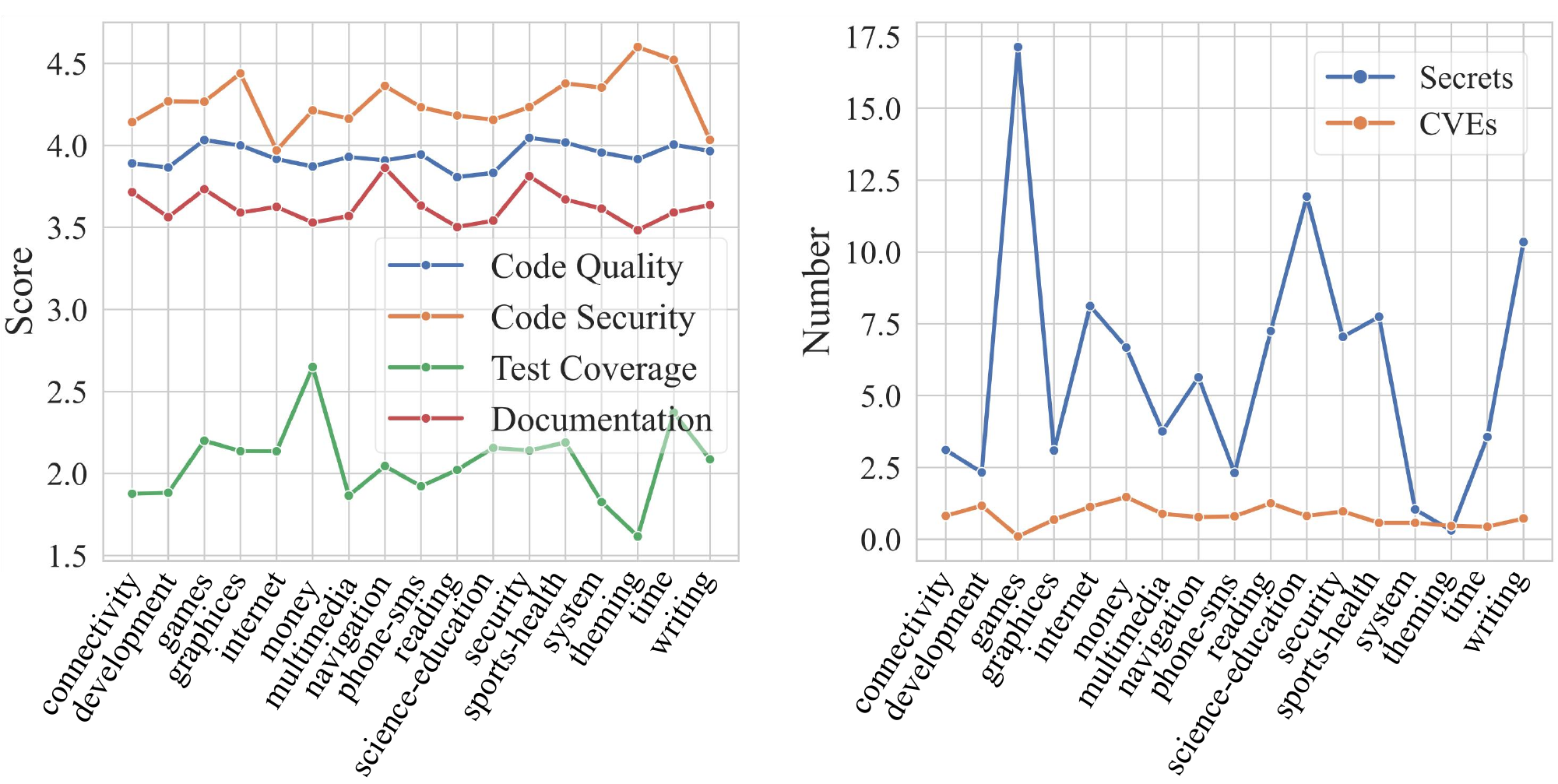} 
\caption{Category Analysis of LLM Scores and Security Exposures. The left subfigure compares category-wise LLM evaluation scores across Code Quality, Code Security, Test Coverage, and Documentation. The right subfigure illustrates the category distribution of CVE and secret exposure counts.}
\label{fig:Category-wise_Analysis} 
\end{figure}

\textbf{Scale and Category Effects.}  
As shown in Figure~\ref{fig:Category-wise_Analysis}, our stratified analysis yields two key insights. 
First, functional categories appear to shape both practices and risks: for example, ``internet'' and ``writing'' apps scored lower on LLM-estimated security, while ``theming'' and ``time'' apps scored higher; ``money'' apps demonstrated stronger testing, whereas ``games'' apps showed the most secret leakage. 
Second, a clear misalignment emerges between LLM-based scores (quality, security, testing, documentation) and empirical outcomes (CVEs, Secrets). 
While LLM evaluations capture plausible surface-level distinctions, they do not consistently reflect real-world vulnerabilities, underscoring both their diagnostic value and their current limitations as security predictors.

\textbf{Insights.}  
Our findings illustrate both the opportunities and the limitations of LLM-based evaluations. 
On the one hand, LLMs offer scalable, expert-like reviews that capture coding style, structural organization, and documentation quality at a granularity rarely feasible with traditional automated tools. 
On the other hand, the weak alignment with empirical outcomes raises concerns about their reliability as standalone risk predictors. 
Future work should focus on improving LLM prompting strategies, developing reasoning scaffolds, and pursuing domain-specific fine-tuning to reduce misclassification. 
Hybrid pipelines that integrate LLM diagnostics with static or dynamic program analysis may further combine contextual reasoning with semantic precision, addressing complementary weaknesses. 
Moreover, large-scale, rigorously validated benchmarks will be essential for calibrating these evaluations across diverse project domains and library ecosystems. 
At the application level, embedding LLM evaluation into IDEs, CI/CD pipelines, or security review dashboards could make these insights directly actionable. 
Ultimately, advancing this line of research may transform LLMs from heuristic evaluators into trustworthy early-warning systems for software vulnerabilities, provided that explainability, reproducibility, and principled evaluation frameworks are in place.

\section{Conclusion}

We presented \sysname, a large-scale benchmark for TPL detection in Android, together with \textit{TPL-Extractor} for precise dependency annotation. Our evaluation of ten representative tools uncovered fundamental trade-offs in efficiency, compatibility, and effectiveness, and highlighted persistent challenges under obfuscation and evolving build pipelines. Beyond TPL detection, our analyses of vulnerabilities, malware, secret leakage, and LLM-based evaluation demonstrate the broader security implications of TPL usage. We hope this work not only advances detection techniques but also fosters a more reliable foundation for future research on Android software supply chain security.

\newpage

\appendix
\section*{Ethical Considerations}

This study complies with established ethical guidelines and adheres to recognized security research best practices. The dataset is constructed exclusively from publicly available data obtained from open-source platforms. We have reviewed the open-source licenses associated with the original data and have restricted the release of research artifacts to academic use under appropriate licensing terms to mitigate potential misuse. The study follows the Menlo Report principles, prioritizing user privacy and security. All identified vulnerabilities or secret exposures are reported through appropriate disclosure channels, and strict confidentiality is maintained throughout the research to ensure that no harm is caused to users or stakeholders.

\section*{Open Science}

In alignment with the USENIX Security Open Science Policy, we commit to releasing all relevant research artifacts upon paper acceptance. This includes the evaluation results of state-of-the-art third-party library detection tools in Android applications, the source code of our \textit{TPL-Extractor} tool, and the constructed dataset. All materials will be made publicly available to the research community via \url{https://anonymous.4open.science/r/TPL-Benchmark-E06A} to facilitate transparency, reproducibility, and further advancements in this domain.

\bibliographystyle{plain}
\bibliography{references}

\begin{thebibliography}{10}

\bibitem{whatsoftwarecompositionanalysissca2023}
What {{Is Software Composition Analysis}} ({{SCA}})?, 2023.
\newblock \url{https://www.mend.io/blog/software-composition-analysis/}.

\bibitem{allatorijavaobfuscatorprofessionaljavaobfuscation}
Allatori {{Java Obfuscator}} - {{Professional Java Obfuscation}}, Accessed: 2025.
\newblock \url{https://allatori.com/}.

\bibitem{androidappsecurityobfuscationdexguard}
Android {{App Security}} and {{Obfuscation}} {\textbar} {{DexGuard}}, Accessed: 2025.
\newblock \url{https://www.guardsquare.com/dexguard}.

\bibitem{appbrain}
Android library statistics, Accessed: 2025.
\newblock \url{https://www.appbrain.com/stats/libraries}.

\bibitem{androidobfuscationjavasecuritydasho2023}
Android {{Obfuscation}} and {{Java Security}} with {{DashO}}, Accessed: 2025.
\newblock \url{https://www.preemptive.com/products/dasho/}.

\bibitem{apachecordova}
Apache {{Cordova}}, Accessed: 2025.
\newblock \url{https://cordova.apache.org/}.

\bibitem{apkpure}
{Download APK to Android using the free online APK downloader in APKPure}, Accessed: 2025.
\newblock \url{https://apkpure.com/cn/}.

\bibitem{enableappoptimizationappquality}
Enable app optimization {\textbar} {{App}} quality, Accessed: 2025.
\newblock \url{https://developer.android.google.cn/topic/performance/app-optimization/enable-app-optimization}.

\bibitem{fdroidfreeopensourceandroidapprepository}
{F-Droid - Free and Open Source Android App Repository}, Accessed: 2025.
\newblock \url{https://f-droid.org/}.

\bibitem{flutterbuildappsanyscreen}
Flutter - {{Build}} apps for any screen, Accessed: 2025.
\newblock \url{https://flutter.dev/}.

\bibitem{gradlebuildtool2024}
Gradle {{Build Tool}}, Accessed: 2025.
\newblock \url{https://gradle.org/}.

\bibitem{javaobfuscatorandroidappoptimizerproguard}
Java {{Obfuscator}} and {{Android App Optimizer}} {\textbar} {{ProGuard}}, Accessed: 2025.
\newblock \url{https://www.guardsquare.com/proguard}.

\bibitem{mavenrepositorysearchbrowseexplore}
Maven {{Repository}}: {{Search}}/{{Browse}}/{{Explore}}, Accessed: 2025.
\newblock \url{https://mvnrepository.com/}.

\bibitem{nvdhome}
Nvd - home, Accessed: 2025.
\newblock \url{https://nvd.nist.gov/}.

\bibitem{reactnative}
React {{Native}}, Accessed: 2025.
\newblock \url{https://reactnative.cn}.

\bibitem{xiaomiappstore}
Xiaomi {{APP Store}}, Accessed: 2025.
\newblock \url{https://m.app.mi.com/}.

\bibitem{aafer2013droidapiminer}
Yousra Aafer, Wenliang Du, and Heng Yin.
\newblock Droidapiminer: Mining api-level features for robust malware detection in android.
\newblock In {\em International conference on security and privacy in communication systems}, pages 86--103. Springer, 2013.

\bibitem{alecci2024androzoo}
Marco Alecci, Pedro Jes{\'u}s~Ruiz Jim{\'e}nez, Kevin Allix, Tegawend{\'e}~F Bissyand{\'e}, and Jacques Klein.
\newblock Androzoo: A retrospective with a glimpse into the future.
\newblock In {\em Proceedings of the 21st International Conference on Mining Software Repositories}, pages 389--393, 2024.

\bibitem{Allix:2016:ACM:2901739.2903508}
Kevin Allix, Tegawend{\'e}~F. Bissyand{\'e}, Jacques Klein, and Yves Le~Traon.
\newblock Androzoo: Collecting millions of android apps for the research community.
\newblock In {\em Proceedings of the 13th International Conference on Mining Software Repositories}, MSR '16, pages 468--471, New York, NY, USA, 2016. ACM.

\bibitem{aonzo2020obfuscapk}
Simone Aonzo, Gabriel~Claudiu Georgiu, Luca Verderame, and Alessio Merlo.
\newblock Obfuscapk: An open-source black-box obfuscation tool for android apps.
\newblock {\em SoftwareX}, 11:100403, 2020.

\bibitem{arzt2014flowdroid}
Steven Arzt, Siegfried Rasthofer, Christian Fritz, Eric Bodden, Alexandre Bartel, Jacques Klein, Yves Le~Traon, Damien Octeau, and Patrick McDaniel.
\newblock Flowdroid: Precise context, flow, field, object-sensitive and lifecycle-aware taint analysis for android apps.
\newblock {\em ACM sigplan notices}, 49(6):259--269, 2014.

\bibitem{avdiienko2015mining}
Vitalii Avdiienko, Konstantin Kuznetsov, Alessandra Gorla, Andreas Zeller, Steven Arzt, Siegfried Rasthofer, and Eric Bodden.
\newblock Mining apps for abnormal usage of sensitive data.
\newblock In {\em 2015 IEEE/ACM 37th IEEE international conference on software engineering}, volume~1, pages 426--436. IEEE, 2015.

\bibitem{backes2016}
Michael Backes, Sven Bugiel, and Erik Derr.
\newblock Reliable {{Third-Party Library Detection}} in {{Android}} and its {{Security Applications}}.
\newblock In {\em Proceedings of the 2016 {{ACM SIGSAC Conference}} on {{Computer}} and {{Communications Security}}}, {{CCS}} '16, pages 356--367, 2016.

\bibitem{bhoraskar2014brahmastra}
Ravi Bhoraskar, Seungyeop Han, Jinseong Jeon, Tanzirul Azim, Shuo Chen, Jaeyeon Jung, Suman Nath, Rui Wang, and David Wetherall.
\newblock Brahmastra: Driving apps to test the security of third-party components.
\newblock In {\em 23rd USENIX Security Symposium (USENIX Security 14)}, pages 1021--1036, 2014.

\bibitem{cohen2013applied}
Jacob Cohen, Patricia Cohen, Stephen~G West, and Leona~S Aiken.
\newblock {\em Applied multiple regression/correlation analysis for the behavioral sciences}.
\newblock Routledge, 2013.

\bibitem{derr2017}
Erik Derr, Sven Bugiel, Sascha Fahl, Yasemin Acar, and Michael Backes.
\newblock Keep me {{Updated}}: {{An Empirical Study}} of {{Third-Party Library Updatability}} on {{Android}}.
\newblock In {\em Proceedings of the 2017 {{ACM SIGSAC Conference}} on {{Computer}} and {{Communications Security}}}, {{CCS}} '17, pages 2187--2200, 2017.

\bibitem{dong2024same}
Zikan Dong, Yanjie Zhao, Tianming Liu, Chao Wang, Guosheng Xu, Guoai Xu, Lin Zhang, and Haoyu Wang.
\newblock Same app, different behaviors: Uncovering device-specific behaviors in android apps.
\newblock In {\em Proceedings of the 39th IEEE/ACM International Conference on Automated Software Engineering}, pages 2099--2109, 2024.

\bibitem{duan2017}
Ruian Duan, Ashish Bijlani, Meng Xu, Taesoo Kim, and Wenke Lee.
\newblock Identifying {{Open-Source License Violation}} and 1-day {{Security Risk}} at {{Large Scale}}.
\newblock In {\em Proceedings of the 2017 {{ACM SIGSAC Conference}} on {{Computer}} and {{Communications Security}}}, {{CCS}} '17, pages 2169--2185, 2017.

\bibitem{enck2011study}
William Enck, Damien Octeau, Patrick~D McDaniel, and Swarat Chaudhuri.
\newblock A study of android application security.
\newblock In {\em USENIX security symposium}, volume~2, pages 1--38, 2011.

\bibitem{geiger2018}
Franz-Xaver Geiger, Ivano Malavolta, Luca Pascarella, Fabio Palomba, Dario Di~Nucci, and Alberto Bacchelli.
\newblock A graph-based dataset of commit history of real-world {{Android}} apps.
\newblock In {\em Proceedings of the 15th {{International Conference}} on {{Mining Software Repositories}}}, {{MSR}} '18, pages 30--33, 2018.

\bibitem{glanz2017}
Leonid Glanz, Sven Amann, Michael Eichberg, Michael Reif, Ben Hermann, Johannes Lerch, and Mira Mezini.
\newblock {{CodeMatch}}: Obfuscation won't conceal your repackaged app.
\newblock In {\em Proceedings of the 2017 11th {{Joint Meeting}} on {{Foundations}} of {{Software Engineering}}}, {{ESEC}}/{{FSE}} 2017, pages 638--648, 2017.

\bibitem{gordon2015information}
Michael~I Gordon, Deokhwan Kim, Jeff~H Perkins, Limei Gilham, Nguyen Nguyen, and Martin~C Rinard.
\newblock Information flow analysis of android applications in droidsafe.
\newblock In {\em NDSS}, volume~15, page 110, 2015.

\bibitem{han2018}
Hongmu Han, Ruixuan Li, and Junwei Tang.
\newblock Identify and {{Inspect Libraries}} in {{Android Applications}}.
\newblock {\em Wireless Personal Communications}, 103(1):491--503, 2018.

\bibitem{he2022}
Qiang He, Bo~Li, Feifei Chen, John Grundy, Xin Xia, and Yun Yang.
\newblock Diversified {{Third-Party Library Prediction}} for {{Mobile App Development}}.
\newblock {\em IEEE Transactions on Software Engineering}, 48(1):150--165, 2022.

\bibitem{he2022msdroid}
Yiling He, Yiping Liu, Lei Wu, Ziqi Yang, Kui Ren, and Zhan Qin.
\newblock Msdroid: Identifying malicious snippets for android malware detection.
\newblock {\em IEEE Transactions on Dependable and Secure Computing}, 20(3):2025--2039, 2022.

\bibitem{huang2023a}
Jianjun Huang, Bo~Xue, Jiasheng Jiang, Wei You, Bin Liang, Jingzheng Wu, and Yanjun Wu.
\newblock Scalably {{Detecting Third-Party Android Libraries With Two-Stage Bloom Filtering}}.
\newblock {\em IEEE Transactions on Software Engineering}, 49(4):2272--2284, 2023.

\bibitem{huang2019}
sirong huang, feifan tao, yuan zhang, and min yang.
\newblock {LibSeeker: A third-party library detection method for Android applications with parameter self-tuning}.
\newblock {\em Journal of Chinese Mini-Micro Computer Systems}, 40(2):332--340, 2019.

\bibitem{jiang2024binaryai}
Ling Jiang, Junwen An, Huihui Huang, Qiyi Tang, Sen Nie, Shi Wu, and Yuqun Zhang.
\newblock Binaryai: Binary software composition analysis via intelligent binary source code matching.
\newblock In {\em Proceedings of the IEEE/ACM 46th International Conference on Software Engineering}, pages 1--13, 2024.

\bibitem{li2025ui}
Jiawei Li, Jiahao Liu, Jian Mao, Jun Zeng, and Zhenkai Liang.
\newblock Ui-ctx: Understanding ui behaviors with code contexts for mobile applications.
\newblock In {\em NDSS}, 2025.

\bibitem{li2015iccta}
Li~Li, Alexandre Bartel, Tegawend{\'e}~F Bissyand{\'e}, Jacques Klein, Yves Le~Traon, Steven Arzt, Siegfried Rasthofer, Eric Bodden, Damien Octeau, and Patrick McDaniel.
\newblock Iccta: Detecting inter-component privacy leaks in android apps.
\newblock In {\em 2015 IEEE/ACM 37th IEEE International Conference on Software Engineering}, volume~1, pages 280--291. IEEE, 2015.

\bibitem{li2019rebooting}
Li~Li, Tegawend{\'e}~F Bissyand{\'e}, and Jacques Klein.
\newblock Rebooting research on detecting repackaged android apps: Literature review and benchmark.
\newblock {\em IEEE Transactions on Software Engineering}, 47(4):676--693, 2019.

\bibitem{li2016}
Li~Li, Tegawend{\'e}~F. Bissyand{\'e}, Jacques Klein, and Yves Le~Traon.
\newblock An {{Investigation}} into the {{Use}} of {{Common Libraries}} in {{Android Apps}}.
\newblock In {\em 2016 {{IEEE}} 23rd {{International Conference}} on {{Software Analysis}}, {{Evolution}}, and {{Reengineering}} ({{SANER}})}, volume~1, pages 403--414, 2016.

\bibitem{li2019identifying}
Li~Li, Tegawend{\'e}~F Bissyand{\'e}, Hao-Yu Wang, and Jacques Klein.
\newblock On identifying and explaining similarities in android apps.
\newblock {\em Journal of Computer Science and Technology}, 34(2):437--455, 2019.

\bibitem{li2019revisiting}
Li~Li, Timoth{\'e}e Riom, Tegawend{\'e}~F Bissyand{\'e}, Haoyu Wang, Jacques Klein, and Le~Traon Yves.
\newblock Revisiting the impact of common libraries for android-related investigations.
\newblock {\em Journal of Systems and Software}, 154:157--175, 2019.

\bibitem{li2018large}
Menghao Li, Pei Wang, Wei Wang, Shuai Wang, Dinghao Wu, Jian Liu, Rui Xue, Wei Huo, and Wei Zou.
\newblock Large-scale third-party library detection in android markets.
\newblock {\em IEEE Transactions on Software Engineering}, 46(9):981--1003, 2018.

\bibitem{li2017c}
Menghao Li, Wei Wang, Pei Wang, Shuai Wang, Dinghao Wu, Jian Liu, Rui Xue, and Wei Huo.
\newblock {{LibD}}: {{Scalable}} and {{Precise Third-Party Library Detection}} in {{Android Markets}}.
\newblock In {\em 2017 {{IEEE}}/{{ACM}} 39th {{International Conference}} on {{Software Engineering}} ({{ICSE}})}, pages 335--346, 2017.

\bibitem{li2023malwukong}
Ningke Li, Shenao Wang, Mingxi Feng, Kailong Wang, Meizhen Wang, and Haoyu Wang.
\newblock Malwukong: Towards fast, accurate, and multilingual detection of malicious code poisoning in oss supply chains.
\newblock In {\em 2023 38th IEEE/ACM International Conference on Automated Software Engineering (ASE)}, pages 1993--2005. IEEE, 2023.

\bibitem{ma2016}
Ziang Ma, Haoyu Wang, Yao Guo, and Xiangqun Chen.
\newblock {{LibRadar}}: Fast and accurate detection of third-party libraries in {{Android}} apps.
\newblock In {\em Proceedings of the 38th {{International Conference}} on {{Software Engineering Companion}}}, {{ICSE}} '16, pages 653--656, 2016.

\bibitem{narayanan2014}
Annamalai Narayanan, Lihui Chen, and Chee~Keong Chan.
\newblock {{AdDetect}}: {{Automated}} detection of {{Android}} ad libraries using semantic analysis.
\newblock In {\em 2014 {{IEEE Ninth International Conference}} on {{Intelligent Sensors}}, {{Sensor Networks}} and {{Information Processing}} ({{ISSNIP}})}, pages 1--6, 2014.

\bibitem{rajpurkar2016squad}
Pranav Rajpurkar, Jian Zhang, Konstantin Lopyrev, and Percy Liang.
\newblock Squad: 100,000+ questions for machine comprehension of text.
\newblock {\em arXiv preprint arXiv:1606.05250}, 2016.

\bibitem{samhi2024}
Jordan Samhi, Tegawend{\'e}~F. Bissyand{\'e}, and Jacques Klein.
\newblock {{AndroLibZoo}}: {{A Reliable Dataset}} of {{Libraries Based}} on {{Software Dependency Analysis}}.
\newblock In {\em 2024 {{IEEE}}/{{ACM}} 21st {{International Conference}} on {{Mining Software Repositories}} ({{MSR}})}, pages 32--36, 2024.

\bibitem{shao2014towards}
Yuru Shao, Xiapu Luo, Chenxiong Qian, Pengfei Zhu, and Lei Zhang.
\newblock Towards a scalable resource-driven approach for detecting repackaged android applications.
\newblock In {\em Proceedings of the 30th Annual Computer Security Applications Conference}, pages 56--65, 2014.

\bibitem{soh2016}
Charlie Soh, Hee~Beng Kuan~Tan, Yauhen~Leanidavich Arnatovich, Annamalai Narayanan, and Lipo Wang.
\newblock {{LibSift}}: {{Automated Detection}} of {{Third-Party Libraries}} in {{Android Applications}}.
\newblock In {\em 2016 23rd {{Asia-Pacific Software Engineering Conference}} ({{APSEC}})}, pages 41--48, 2016.

\bibitem{tang2020a}
Wei Tang, Ping Luo, Jialiang Fu, and Dan Zhang.
\newblock {{LibDX}}: {{A Cross-Platform}} and {{Accurate System}} to {{Detect Third-Party Libraries}} in {{Binary Code}}.
\newblock In {\em 2020 {{IEEE}} 27th {{International Conference}} on {{Software Analysis}}, {{Evolution}} and {{Reengineering}} ({{SANER}})}, pages 104--115, 2020.

\bibitem{tang2019}
Zhushou Tang, Minhui Xue, Guozhu Meng, Chengguo Ying, Yugeng Liu, Jianan He, Haojin Zhu, and Yang Liu.
\newblock Securing android applications via edge assistant third-party library detection.
\newblock {\em Computers \& Security}, 80:257--272, 2019.

\bibitem{team2024qwen2}
Qwen Team.
\newblock Qwen2 technical report.
\newblock {\em arXiv preprint arXiv:2407.10671}, 2024.

\bibitem{umayya2024comex}
Zeya Umayya, Dhruv Malik, Arpit Nandi, Akshat Kumar, Sareena Karapoola, and Sambuddho Chakravarty.
\newblock Comex: Deeply observing application behavior on real android devices.
\newblock In {\em Proceedings of the 17th Cyber Security Experimentation and Test Workshop}, pages 100--109, 2024.

\bibitem{wang2015a}
Haoyu Wang, Yao Guo, Ziang Ma, and Xiangqun Chen.
\newblock {{WuKong}}: A scalable and accurate two-phase approach to {{Android}} app clone detection.
\newblock In {\em Proceedings of the 2015 {{International Symposium}} on {{Software Testing}} and {{Analysis}}}, {{ISSTA}} 2015, pages 71--82, 2015.

\bibitem{wang2019understanding}
Haoyu Wang, Hao Li, and Yao Guo.
\newblock Understanding the evolution of mobile app ecosystems: A longitudinal measurement study of google play.
\newblock In {\em The World Wide Web Conference}, pages 1988--1999, 2019.

\bibitem{wang2018beyond}
Haoyu Wang, Zhe Liu, Jingyue Liang, Narseo Vallina-Rodriguez, Yao Guo, Li~Li, Juan Tapiador, Jingcun Cao, and Guoai Xu.
\newblock Beyond google play: A large-scale comparative study of chinese android app markets.
\newblock In {\em Proceedings of the Internet Measurement Conference 2018}, pages 293--307, 2018.

\bibitem{wang2018c}
Yan Wang, Haowei Wu, Hailong Zhang, and Atanas Rountev.
\newblock {{ORLIS}}: Obfuscation-resilient library detection for {{Android}}.
\newblock In {\em Proceedings of the 5th {{International Conference}} on {{Mobile Software Engineering}} and {{Systems}}}, {{MOBILESoft}} '18, pages 13--23, 2018.

\bibitem{wei2018amandroid}
Fengguo Wei, Sankardas Roy, Xinming Ou, and Robby.
\newblock Amandroid: A precise and general inter-component data flow analysis framework for security vetting of android apps.
\newblock {\em ACM Transactions on Privacy and Security (TOPS)}, 21(3):1--32, 2018.

\bibitem{willmott2005advantages}
Cort~J Willmott and Kenji Matsuura.
\newblock Advantages of the mean absolute error (mae) over the root mean square error (rmse) in assessing average model performance.
\newblock {\em Climate research}, 30(1):79--82, 2005.

\bibitem{wu2024identifying}
Susheng Wu, Wenyan Song, Kaifeng Huang, Bihuan Chen, and Xin Peng.
\newblock Identifying affected libraries and their ecosystems for open source software vulnerabilities.
\newblock In {\em Proceedings of the IEEE/ACM 46th International Conference on Software Engineering}, pages 1--12, 2024.

\bibitem{wu2023a}
Yafei Wu, Cong Sun, Dongrui Zeng, Gang Tan, Siqi Ma, and Peicheng Wang.
\newblock Libscan: Towards more precise third-party library identification for android applications.
\newblock In {\em 32nd {{USENIX Security Symposium}} ({{USENIX Security}} 23)}, pages 3385--3402, 2023.

\bibitem{wu2019malscan}
Yueming Wu, Xiaodi Li, Deqing Zou, Wei Yang, Xin Zhang, and Hai Jin.
\newblock Malscan: Fast market-wide mobile malware scanning by social-network centrality analysis.
\newblock In {\em 2019 34th IEEE/ACM International Conference on Automated Software Engineering (ASE)}, pages 139--150. IEEE, 2019.

\bibitem{xie2023a}
Zifan Xie, Ming Wen, Haoxiang Jia, Xiaochen Guo, Xiaotong Huang, Deqing Zou, and Hai Jin.
\newblock Precise and {{Efficient Patch Presence Test}} for {{Android Applications}} against {{Code Obfuscation}}.
\newblock In {\em Proceedings of the 32nd {{ACM SIGSOFT International Symposium}} on {{Software Testing}} and {{Analysis}}}, {{ISSTA}} 2023, pages 347--359, 2023.

\bibitem{xie2024b}
Zifan Xie, Ming Wen, Tinghan Li, Yiding Zhu, Qinsheng Hou, and Hai Jin.
\newblock How {{Does Code Optimization Impact Third-party Library Detection}} for {{Android Applications}}?
\newblock In {\em Proceedings of the 39th {{IEEE}}/{{ACM International Conference}} on {{Automated Software Engineering}}}, {{ASE}} '24, pages 1919--1931, 2024.

\bibitem{xu2020}
Jian Xu and Qianting Yuan.
\newblock {{LibRoad}}: {{Rapid}}, {{Online}}, and {{Accurate Detection}} of {{TPLs}} on {{Android}}.
\newblock {\em IEEE Transactions on Mobile Computing}, pages 1--1, 2020.

\bibitem{xujian2023}
Jian XU and QianTing YUAN.
\newblock {{LibPass}}: {{Third-party Library Detection Method Based}} on {{Package Structure}} and {{Signature}}.
\newblock {\em Journal of Software}, 35(6):2880--2902, 2023.

\bibitem{xu2024enhancing}
Shangzhi Xu, Jialiang Dong, Weiting Cai, Juanru Li, Arash Shaghaghi, Nan Sun, and Siqi Ma.
\newblock Enhancing security in third-party library reuse--comprehensive detection of 1-day vulnerability through code patch analysis.
\newblock {\em arXiv preprint arXiv:2411.19648}, 2024.

\bibitem{xu2017neural}
Xiaojun Xu, Chang Liu, Qian Feng, Heng Yin, Le~Song, and Dawn Song.
\newblock Neural network-based graph embedding for cross-platform binary code similarity detection.
\newblock In {\em Proceedings of the 2017 ACM SIGSAC conference on computer and communications security}, pages 363--376, 2017.

\bibitem{yu2020order}
Zeping Yu, Rui Cao, Qiyi Tang, Sen Nie, Junzhou Huang, and Shi Wu.
\newblock Order matters: Semantic-aware neural networks for binary code similarity detection.
\newblock In {\em Proceedings of the AAAI conference on artificial intelligence}, volume~34, pages 1145--1152, 2020.

\bibitem{zhan2021a}
Xian Zhan, Lingling Fan, Sen Chen, Feng We, Tianming Liu, Xiapu Luo, and Yang Liu.
\newblock {{ATVHunter}}: {{Reliable Version Detection}} of {{Third-Party Libraries}} for {{Vulnerability Identification}} in {{Android Applications}}.
\newblock In {\em 2021 {{IEEE}}/{{ACM}} 43rd {{International Conference}} on {{Software Engineering}} ({{ICSE}})}, pages 1695--1707, 2021.

\bibitem{zhan2019comparative}
Xian Zhan, Tao Zhang, and Yutian Tang.
\newblock A comparative study of android repackaged apps detection techniques.
\newblock In {\em 2019 IEEE 26th International Conference on Software Analysis, Evolution and Reengineering (SANER)}, pages 321--331. IEEE, 2019.

\bibitem{zhang2018re}
Chengpeng Zhang, Haoyu Wang, Ran Wang, Yao Guo, and Guoai Xu.
\newblock Re-checking app behavior against app description in the context of third-party libraries.
\newblock In {\em SEKE}, pages 665--664, 2018.

\bibitem{zhang2019}
Jiexin Zhang, Alastair~R. Beresford, and Stephan~A. Kollmann.
\newblock {{LibID}}: Reliable identification of obfuscated third-party {{Android}} libraries.
\newblock In {\em Proceedings of the 28th {{ACM SIGSOFT International Symposium}} on {{Software Testing}} and {{Analysis}}}, {{ISSTA}} 2019, pages 55--65, 2019.

\bibitem{zhang2018b}
Yuan Zhang, Jiarun Dai, Xiaohan Zhang, Sirong Huang, Zhemin Yang, Min Yang, and Hao Chen.
\newblock Detecting third-party libraries in {{Android}} applications with high precision and recall.
\newblock In {\em 2018 {{IEEE}} 25th {{International Conference}} on {{Software Analysis}}, {{Evolution}} and {{Reengineering}} ({{SANER}})}, pages 141--152, 2018.

\bibitem{zhang2020a}
Zicheng Zhang, Wenrui Diao, Chengyu Hu, Shanqing Guo, Chaoshun Zuo, and Li~Li.
\newblock An empirical study of potentially malicious third-party libraries in {{Android}} apps.
\newblock In {\em Proceedings of the 13th {{ACM Conference}} on {{Security}} and {{Privacy}} in {{Wireless}} and {{Mobile Networks}}}, {{WiSec}} '20, pages 144--154, 2020.

\bibitem{zhao2023a}
Lida Zhao, Sen Chen, Zhengzi Xu, Chengwei Liu, Lyuye Zhang, Jiahui Wu, Jun Sun, and Yang Liu.
\newblock Software {{Composition Analysis}} for {{Vulnerability Detection}}: {{An Empirical Study}} on {{Java Projects}}.
\newblock In {\em Proceedings of the 31st {{ACM Joint European Software Engineering Conference}} and {{Symposium}} on the {{Foundations}} of {{Software Engineering}}}, {{ESEC}}/{{FSE}} 2023, pages 960--972, 2023.

\bibitem{zhauniarovich2014fsquadra}
Yury Zhauniarovich, Olga Gadyatskaya, Bruno Crispo, Francesco La~Spina, and Ermanno Moser.
\newblock Fsquadra: Fast detection of repackaged applications.
\newblock In {\em IFIP Annual Conference on Data and Applications Security and Privacy}, pages 130--145. Springer, 2014.

\bibitem{zheng2024towards}
Xinyi Zheng, Chen Wei, Shenao Wang, Yanjie Zhao, Peiming Gao, Yuanchao Zhang, Kailong Wang, and Haoyu Wang.
\newblock Towards robust detection of open source software supply chain poisoning attacks in industry environments.
\newblock In {\em Proceedings of the 39th IEEE/ACM international conference on automated software engineering}, pages 1990--2001, 2024.

\bibitem{zhou2022uncovering}
Hao Zhou, Haoyu Wang, Xiapu Luo, Ting Chen, Yajin Zhou, and Ting Wang.
\newblock Uncovering cross-context inconsistent access control enforcement in android.
\newblock In {\em The 2022 Network and Distributed System Security Symposium (NDSS'22)}, 2022.

\bibitem{zhou2025hey}
Jiawei Zhou, Zidong Zhang, Lingyun Ying, Huajun Chai, Jiuxin Cao, and Haixin Duan.
\newblock Hey, your secrets leaked! detecting and characterizing secret leakage in the wild.
\newblock In {\em 2025 IEEE Symposium on Security and Privacy (SP)}, pages 449--467. IEEE, 2025.

\end{thebibliography}

\appendices
\begin{appendices}
\section{Reasons for Multiple APKs in an Open-Source Repository}
\label{sec:A}
In open-source Android repositories, it is common to encounter multiple APK files associated with a single project. This phenomenon often puzzles users and researchers who expect one repository to correspond to one application binary. In practice, however, the diversity of development, testing, and distribution requirements leads developers to generate and maintain multiple APKs within the same repository. The following summarizes the major reasons behind this practice.
\begin{itemize}
    \item To support various devices, developers typically generate separate APK files for different CPU architectures, such as ARM, x86, x86\_64, and ARM64. \item Applications may generate the following types of APKs based on their development stages: Debug version (Debug APK), Release version (Release APK), Beta version (Beta APK), Nightly build version (Nightly APK), Aligned version (Aligned APK), Signed version (Signed APK). 
    \item Developers often define different product flavors to meet diverse requirements, such as: Free, paid, or premium (Pro) versions. 
    \item Region-specific versions (e.g., Firefox Focus and Firefox Klar). 
    \item APKs with or without dependency on Google Mobile Services (GMS), such as ``GMS'' and ``FOSS/Floss'' versions. 
    \item Some repositories store old versions of APKs for reference or archiving purposes. 
\end{itemize}

\section{Keywords for Classifying Third-Party Library Dependencies Based on Their Inclusion in the Application Package}
\label{sec:B}

In Android Gradle builds, TPLs can be declared with a variety of dependency keywords. 
However, not all of them result in the corresponding code being packaged into the final APK. 
Correctly distinguishing these keywords is essential for reliable TPL detection, since mistakenly including or excluding a dependency keyword may lead to false positives or false negatives in library identification. 
We therefore categorize the commonly observed keywords into three groups, as shown below.

\noindent
Keywords whose dependencies are included in the APK:
\small
\begin{lstlisting}
withAnalyticsImplementation, natives, api, compile,
androidImplementation, implementation,
releaseImplementation, releaseCompile,
coreLibraryDesugaring, etc.
\end{lstlisting}
\noindent
\normalsize
Keywords whose dependencies are not included in the APK: 
\small
\begin{lstlisting}
testImplementation, androidTestImplementation, Kapt,
compileOnly, debugImplementation, androidTestApt,
annotationProcessor, testApi, apt, jnaForTest,
retrolambdaConfig, detektPlugins, debugcompile,
androidTestApi, androidTestUtil, errorprone, ksp,
kaptAndroidTest, testAnnotationProcessor, ktlint,
androidTestAnnotationProcessor, etc. 
\end{lstlisting}
\noindent
\normalsize
Keywords dynamically determined by flavorDimensions, buildType and productFlavors: 
\small
\begin{lstlisting}
PlayStoreImplementation, nightlyImplementation,
gplayImplementation, playImplementation,
largeImplementation, playstoreImplementation,
amazonImplementation, githubImplementation,
pureImplementation, prodImplementation,
betaImplementation, alphaImplementation,
devImplementation, customImplementation, 
appengineSdk, etc.
\end{lstlisting}

\normalsize
\section{Representative Dependency Declaration Patterns}
\label{sec:C}

Table~\ref{tab:dep_patterns} summarizes illustrative instances of the eight common dependency declaration patterns.  
We abbreviate overly long coordinates with ellipses (\texttt{...}) for readability.  
Gradle officially mentions two dependency declaration styles: string notation and map notation. However, our case studies reveal additional practices in real-world projects. In particular, dependency coordinates are frequently constructed via variable concatenation, with values sourced from both build scripts and property files. Although Gradle recommends \texttt{TOML} for centralized dependency management, we also observe the use of custom \texttt{Gradle/Kt} scripts. Developers further employ techniques such as declaring multiple dependencies in a single statement and managing version alignment through BOM declarations.

\begin{table*}[t]
\centering
\scriptsize
\caption{Instances of dependency declaration patterns in Gradle-based Android projects}
\begin{tabularx}{0.97\textwidth}{p{2.9cm} X p{5.0cm}}
\toprule
\textbf{Pattern Type} & \textbf{Example Snippet} & \textbf{Key Characteristic} \\
\midrule
String notation 
& \texttt{compile 'com.android.support:transition:25.1.0'} 
& Combine group, name, and version in a single string. \\
\midrule
Map notation 
& \texttt{compile group:'org.hibernate',name:'hibernate',version:'3.0.5'} 
& Specify each part of the coordinates separately. \\

\midrule
Variable-based declaration 
& \texttt{implementation("\$okhttpGroup:okhttp:\$okhttpVersion")}  
\newline \texttt{implementation "androidx.core:core-ktx:" + versions.core} 
\newline \texttt{compile "androidx.core:core-ktx:\$rootProject.coreVersion"} 
\newline \texttt{compile "androidx.core:core-ktx:\$rootProject.ext.versions.core"} 
\newline \texttt{compile "androidx.activity:activity:\$\{project.ACTIVITY\_VERSION\}"} 
\newline \texttt{compile "androidx.activity:activity:\${project.versions.activity}"} 
\newline \texttt{compile rootProject.ext.dependencies["com.android.support:design"]}
& Coordinates built via variable interpolation, string concatenation, or property references (\$var, \$rootProject, \${project.xxx}, ext). \\
 
\midrule
Multi-library declaration 
& \texttt{implementation("...:appcompat-v7:\$support\_lib\_version", "...:aspects:1.0.0@aar", "...:butterknife:\$butterknife\_version", "...:butterknife-aspects:1.0.0")} 
& Declare multiple libraries in a single statement. \\
\midrule
Version catalog (\texttt{TOML}) 
& \texttt{implementation libs.bundles.android} 
& Defines library versions and aliases in a central file. \\
\midrule
Custom \texttt{Gradle/Kt} script 
& \texttt{implementation deps.kotlin.stdlib.jdk8} 
& Dependency supplied by custom script object. \\
\midrule
Bill of Materials (BOM) 
& \texttt{implementation(platform("com.squareup.okhttp3:okhttp-bom:4.11.0"))} \newline \texttt{implementation 'com.squareup.okhttp3:okhttp'} \newline \texttt{implementation 'com.squareup.okhttp3:logging-interceptor'} 
& BOM supplies versions; dependent libs omit explicit version. \\
\midrule
BOM via catalog reference 
& \texttt{gplayImplementation platform(libs.google.firebaseBom)} \newline \texttt{gplayImplementation(libs.google.messaging)} 
& BOM imported through the version catalog handle. \\
\bottomrule
\end{tabularx}
\label{tab:dep_patterns}
\end{table*}

\normalsize
\section{Examples of Variable Assignment Modes}
\label{sec:D}

\noindent
We provide concrete examples of how dependency-related variables are assigned and organized in Gradle-based Android projects. 
These snippets illustrate five representative modes observed in the wild. 
Such patterns demonstrate the heterogeneity of variable management across projects, 
ranging from flat declarations to deeply nested objects and standardized version catalogs.
\small
\begin{lstlisting}
(1) Direct assignment
ext.anko_version = '0.10.0'
(2) ext block
ext { playServicesVersion = '17.0.0' }
(3) Variables map
ext.versions = [play: '11.6.2']
ext.deps = [material:'com.google.android.material:material:1.0.0']
(4) Nested structure
ext { libVersions = [android: [espresso:'3.0.2']] }
(5) Version catalog (TOML)
[versions]
core = "1.0.0"
[libraries]
core-ktx = { module = "androidx.core:core-ktx", version.ref = "core" }
[bundles]
core = ["core-ktx"]
\end{lstlisting}

\normalsize
\section{The Algorithm Procedure of \textit{TPL-Extractor}}
\label{sec:E}

\noindent
Algorithm~\ref{alg:tpl_extractor} presents the detailed procedure of \textit{TPL-Extractor}, 
our framework for extracting TPL dependencies from Android projects. 
Given a project $p$, the algorithm systematically parses project- and module-level Gradle files, \texttt{TOML}-based version catalogs, and global property files, resolves variable definitions, 
and constructs normalized dependency triplets $T_p$. 
The process ensures accurate extraction even in complex modular projects with nested or cross-module variable references.

\begin{algorithm}[h]
\footnotesize
\caption{\textit{TPL-Extractor} Framework for Extracting Third-Party Library Dependencies}
\label{alg:tpl_extractor}
\begin{algorithmic}[1]
\Require Android project $p$
\Ensure Extracted third-party library triplets $T_p$

\Statex \textbf{Step 1: Identify Root Directory}
\State Locate root directory $R_p$ containing: 
\State \hspace{1em} \texttt{build.gradle(.kts)}, \texttt{settings.gradle(.kts)}, etc.

\Statex \textbf{Step 2: Extract Global Variables}
\State Parse \texttt{gradle.properties} for global vars $K_p$
\State Parse project-level \texttt{build.gradle(.kts)} for shared configs

\Statex \textbf{Step 3: Parse TOML Files}
\If{\texttt{libs.version.toml} or \texttt{libs.toml} exists in $R_p$}
    \State Extract dependency entries into $\mathcal{D}^{\mathrm{toml}}_p$
\EndIf

\Statex \textbf{Step 4: Analyze Module Configs}
\ForAll{module $m \in A_p$}
    \State Extract local vars $K_{p,m}$ and dependency declarations $\mathcal{D}_{p,m}$
    \State Classify project type by $|A_p|$:
    \If{$|A_p| = 1$}
        \State Merge $K_p$ and $K_{p,m}$
    \ElsIf{$|A_p| = 2$}
        \State Separate $K_p$ and $K_{p,m}$
    \ElsIf{$|A_p| \geq 3$}
        \State \textbf{Main module:} direct APK dependencies
        \State \textbf{Submodules:} indirect dependencies
    \EndIf
\EndFor

\Statex \textbf{Step 5: Build Dependency Graph}
\State Construct $G_p = (A_p, E_p)$ from module references
\State Apply topological sort on $A_p$ to get build order

\Statex \textbf{Step 6: Identify Main Module}
\State Identify $m_{\mathrm{main}} \in A_p$ via tags like \texttt{com.android.application},etc. 

\Statex \textbf{Step 7: Aggregate and Clean Data}
\State Aggregate reachable module declarations: 
\State \hspace{1em} $\widehat{\mathcal{D}}_p = \{(m,d) \mid m\in A_p,\, d\in \mathcal{D}_{p,m}\}$
\State Parse and normalize into raw triplets: $T_p^{\mathrm{raw}} = \mathcal{N}(\widehat{\mathcal{D}}_p, K_p, \{K_{p,m}\})$
\State Deduplicate: $T_p^{\mathrm{uniq}} = \Delta(T_p^{\mathrm{raw}})$
\State Resolve version conflicts: $T_p = \rho(T_p^{\mathrm{uniq}})$

\State \Return $T_p$
\end{algorithmic}
\end{algorithm}

\section{Names of Custom Configuration Files}
\label{sec:F}

\noindent
Custom Gradle and Kotlin files encountered in Android projects may contain variable declarations or direct dependency triplets. These scripts are irregularly named. We summarize the custom file names encountered in our research as follows:

\small
\begin{lstlisting}
constants.gradle, versions.gradle, config.gradle,
dependencies_groups.gradle, dependencies.gradle,
dependency-versions.gradle, variables.gradle,
deps.gradle, sdkVersion.gradle, standalone.gradle,
root_all_projects_ext.gradle, dependencies.kt,
dependencies.gradle.kts, Deps.kt, deps.kt, Vers.kt,
Versions.kt.
\end{lstlisting}

\normalsize
\section{Parameter Settings and Evaluation Details}
\label{sec:G}

For completeness, we summarize the parameter configurations and evaluation details of the ten third-party library detection tools used in our experiments:

\textit{Libradar.} Libradar reports both package structures of libraries in applications (e.g., \texttt{com/a/a}) and known standard library structures(e.g., \texttt{com/support/design}). We extract all package structures from our TPL database and match them against Libradar's outputs to evaluate its detection results.

\textit{LibD.} LibD produces only package structures of libraries in applications, which are often heavily affected by obfuscated names (e.g., \texttt{f/e/a}, \texttt{t}, \texttt{E5}). Such outputs are excluded from evaluation due to their ambiguity.

\textit{LibScout.} Package similarity threshold set to $0.5$, library similarity threshold set to $0.6$.

\textit{LibPecker.} Same thresholds as LibScout, i.e., $0.5$ (package) and $0.6$ (library).

\textit{Orlis.} Detection criterion: an application is considered to include a library if at least $60\%$ of the library's classes are present.

\textit{LibID.} Parameters configured as $\Gamma_1 = 0.8$, $\Gamma_2 = 0.1$.

\textit{LibScan.} Class similarity threshold $\theta_1 = 0.7$, library confidence score threshold $\theta_2 = 0.85$.

\textit{LibHunter.} Method-level threshold $T_1 = 0.75$, TPL-level threshold $T_2 = 0.2$.

\textit{LibLoom.} Library similarity threshold set to $0.6$, with other parameters following the original paper.

All other hyperparameters and preprocessing follow the respective default settings as reported in the original publications.

\section{Limitations and Discussion}
\label{sec:H}

Our work inevitably carries certain limitations that should be acknowledged with candor. 
First, the dataset construction, while carefully curated, is primarily based on open-source projects (e.g., F-Droid and GitHub) and therefore may not fully represent the characteristics of proprietary applications. 
Source-APK alignment is not always perfect, as some repositories are incomplete or distribute dependencies across multiple submodule repositories, and some dependencies may be difficult to resolve due to deprecated repositories or dynamic versioning. 
Although we attempted to mitigate these issues through manual supplementation and multi-source validation, the resulting ground truth should be interpreted as a practical approximation rather than an absolute standard.  

Second, our findings are influenced by the Android ecosystem and the external resources we rely on (e.g., VirusTotal, Gitleaks, NVD), which may contain their own gaps or inconsistencies. 
While such imperfections are common in large-scale empirical studies, they remind us that our conclusions should be viewed as indicative rather than definitive. 
However, we believe that these limitations do not undermine the central insights of this paper, namely, the confounding impact of ubiquitous libraries on malware detection, risks of secret leakage in open projects, and the importance of accurate TPL identification. 
Future work could extend the dataset to more diverse sources, explore automated handling of incomplete dependencies, and investigate cross-platform ecosystems, thereby further strengthening the generality of the results. 

\end{appendices}

\end{document}